\documentclass[twocolumn]{aastex63}

\usepackage[figuresright]{rotating}
\usepackage{kotex}

\graphicspath{{./}{figures/}}
\accepted{for publication in ApJS}

\shorttitle{Subaru/FOCAS MOS Observation of M87 Globular Clusters}
\shortauthors{Kim et al.}

\begin{document}

\title{Nonlinear Color$-$Metallicity Relations of Globular Clusters. X. Subaru/FOCAS Multi-object Spectroscopy of M87 Globular Clusters}

\correspondingauthor{Suk-Jin Yoon}
\email{sjyoon0691@yonsei.ac.kr}

\author[0000-0001-6574-675X]{Sooyoung Kim}
\affil{Korea Astronomy and Space Science Institute, Daejeon 34055, Republic of Korea}

\author[0000-0002-1842-4325]{Suk-Jin Yoon}
\affil{Department of Astronomy, Yonsei University, Seoul 03722, Republic of Korea}
\affil{Center for Galaxy Evolution Research, Yonsei University, Seoul 03722, Republic of Korea}

\author[0000-0002-7957-3877]{Sang-Yoon Lee}
\affil{Center for Galaxy Evolution Research, Yonsei University, Seoul 03722, Republic of Korea}

\author[0000-0001-6812-4542]{Chul Chung}
\affil{Department of Astronomy, Yonsei University, Seoul 03722, Republic of Korea}
\affil{Center for Galaxy Evolution Research, Yonsei University, Seoul 03722, Republic of Korea}

\author[0000-0001-8368-0221]{Sangmo Tony Sohn}
\affil{Space Telescope Science Institute, 3700 San Martin Drive, Baltimore, MD 21237, U.S.A.}

\begin{abstract}

We obtained spectra of some 140 globular clusters (GCs) associated with the Virgo central cD galaxy M87 with the Subaru/FOCAS MOS mode.
The fundamental properties of GCs such as age, metallicity and $\alpha$-element abundance are investigated by using simple stellar population models.
It is confirmed that the majority of M87 GCs are as old as, more metal-rich than, and more enhanced in $\alpha$-elements than the Milky Way GCs.
Our high-quality, homogeneous dataset enables us to test the theoretical prediction of {\it inflected} color$-$metallicity relations (CMRs). 
The nonlinear-CMR hypothesis entails an alternative explanation for the widely observed GC color bimodality, in which even a unimodal metallicity spread yields a bimodal color distribution by virtue of nonlinear metallicity-to-color conversion.
The newly derived CMRs of old, high-signal-to-noise-ratio GCs in M87 (the $V-I$ CMR of 83 GCs and the $M-T2$ CMR of 78 GCs) corroborate the presence of the significant inflection. 
Furthermore, from a combined catalog with the previous study on M87 GC spectroscopy, we find that a total of 185 old GCs exhibit a broad, unimodal metallicity distribution.
The results corroborate the nonlinear-CMR interpretation of the GC color bimodality, shedding further light on theories of galaxy formation.

\end{abstract}


\section{Introduction}

Globular clusters (GCs) are an old simple stellar population (SSP) with a small internal dispersion in age and chemical composition, and provide powerful clues to help understand the formation histories of their host galaxies.
The exciting discovery of GC color bimodality in many luminous early-type galaxies \cite[e.g.,][]{geisler96, kundu01, larsen01, peng04, harris06,
peng06, lee08, jordan09, sinnott10, liu11, blakeslee12, brodie12, pota13, kartha14, richtler15, cho16, kartha2016, caso2017, harris2017, choksi2018, forbes2018, ennis2019, debortoli20} has been commonly accepted as the presence of two GC subpopulations in a galaxy.
The implications of two distinct populations of GCs in the context of galaxy and GC formation have been widely explored and led to several scenarios \cite[e.g.,][]{ashman92, cote98, forbes97, lee10}.

An alternative, the contrary claim has been suggested by \citet[][hereafter Paper I]{yoon06} where the existence of two distinct GC subsystems is not a necessary assumption to explain the GC color bimodality (see also \citet{richtler06}). 
The conventional scenarios assert that at a fixed age, GC metallicities have a linear relation with color. 
However, via the stellar population synthesis technique, Paper I presented a significant inflection along the color$-$metallicity relations (CMRs). 
In their scenario, the nonlinear CMRs between intrinsic metallicity and its proxy, colors, holds the key to the color bimodality phenomenon. 
Paper I demonstrated that conversions to colors from the unimodal metallicity distribution of old GCs ($>$\,10 Gyr) can be naturally achieved through nonlinear CMRs. 
Nonlinear relations between [Fe/H] and colors are also reported observationally \citep{peng06, blakeslee12, chies12, cho16}.

The nonlinear-CMR scenario was scrutinized in a subsequent series of papers \cite[][hereafter Papers II, III, IV, V, VI, VII, VIII, and IX]{yoon11a, yoon11b, yoon13, kim13, chung2016, kim17, sylee2019, sylee2020}. 
In Papers II and IV, the shapes of CMRs according to different colors were examined for the GC systems of M87 and M84, respectively.
They showed that the $u$-band-related colors in particular, are useful diagnostics of the underlying metallicity distribution function.
Different combinations of $ugz$ bands from Hubble Space Telescope (HST) multiband photometry yielded histograms of varying morphologies, which is consistent with the nonlinear-CMR assumption.
Paper III focused on the analysis of GC metallicities and their link to the host galaxy.
It was shown that when transformed through nonlinear CMRs, GCs and halo field stars share similar metallicity distribution shapes. This suggests a connected origin of GCs and halo field stars, contrary to previous views.
Papers V and VII examined the spectra of old, spheroidal GCs in the M31 and NGC 5128, respectively.
The observed distributions of high-quality spectroscopic absorption-line indices of M31 GCs \citep{caldwell11} and NGC 5128 GCs \citep{beasley08, woodley10} were reproduced successfully using the nonlinear metallicity-to-index transformation, exactly analogous to the view that the nonlinear metallicity-to-color transformation is responsible for the photometric color bimodality of GCs.
A further theoretical spectroscopic study, Paper VI, showed that the predictions from the Yonsei Evolutionary Population Synthesis (YEPS) models \citep{chung13, chung17} exhibit nonlinearity in the Ca II triplet ($CaT$; 8498, 8542, and 8662 \AA) index versus metallicity plane. 
The $CaT$ is a well-known metallicity indicator for stellar populations and is used to derive metallicities of extragalactic GCs assuming a linear relation between $CaT$ and metallicity \cite[e.g.,][]{foster10,brodie12, usher15}.
They also showed that the conversion via nonlinear $CaT-$metallicity from a unimodal metallicity distribution reproduces the observed $CaT$ distribution well.
More recently, two theoretical photometric studies also reinforced the nonlinearity scenario for the GC color bimodality.
Paper VIII showed that the individual color distributions of 78 GC systems in early-type galaxies from the ACS Virgo and Fornax Cluster Surveys are correctly reproduced based on the nonlinear-CMR scenario.
Paper IX showed that the difference between blue and red GCs in M49, M60, M87, and NGC 1399 in terms of the radial surface number density profile arises naturally from the nonlinear metallicity-to-color conversion together with the observed radial metallicity gradient of the GC systems.

This paper is the 10th in the nonlinear-CMR series and adopts the same line of analysis in exploring and determining  more precise forms of GC CMR. 
We use a homogeneous set of high-quality M87 GC spectra with wide metallicity ranges, obtained with the Subaru/FOCAS MOS mode. 
M87 is a cD galaxy at the center of the Virgo cluster harboring a population of $\sim10^4$ GCs.
The previous spectroscopic study of its GC system \cite[][hereafter C98]{cohen98} reported ages and metallicities for $\sim$\,100 GCs. 
Our sample GCs are located similarly within 10$'$ of galactic radius and have nine GCs in common with the C98 catalog.
C98 spectroscopic GCs are brighter than ours by $0.5-1.0$ mag in the $V$ band.
Recently, \citet{Villaume19} reported a new metallicity measurement on M87 GCs via full-spectrum SSP model fitting using the C98 GC in which they obtained systematically higher metallicities compared to the previous findings on M87 (see the discussion).

The paper is outlined as follows. 
Section 2 presents the observation and reduction of the data used in the study. 
Section 3 describes the measurements of radial velocities and selection of member GCs.
Section 4 provides the measurement of Lick/IDS absorption-line indices \cite[][hereafter W94 and W97, respectively]{worthey94, worthey97}, followed by the inferred age, metallicity, and $\alpha$-element abundance of the M87 GCs in Section 5.
In Section 6, we present the derived CMRs of the clusters. 
The implications of our findings are discussed in Section 7.

\section{Observations and Data Reduction}

\begin{figure}[h!]
\begin{center}
\includegraphics[width=13.5cm]{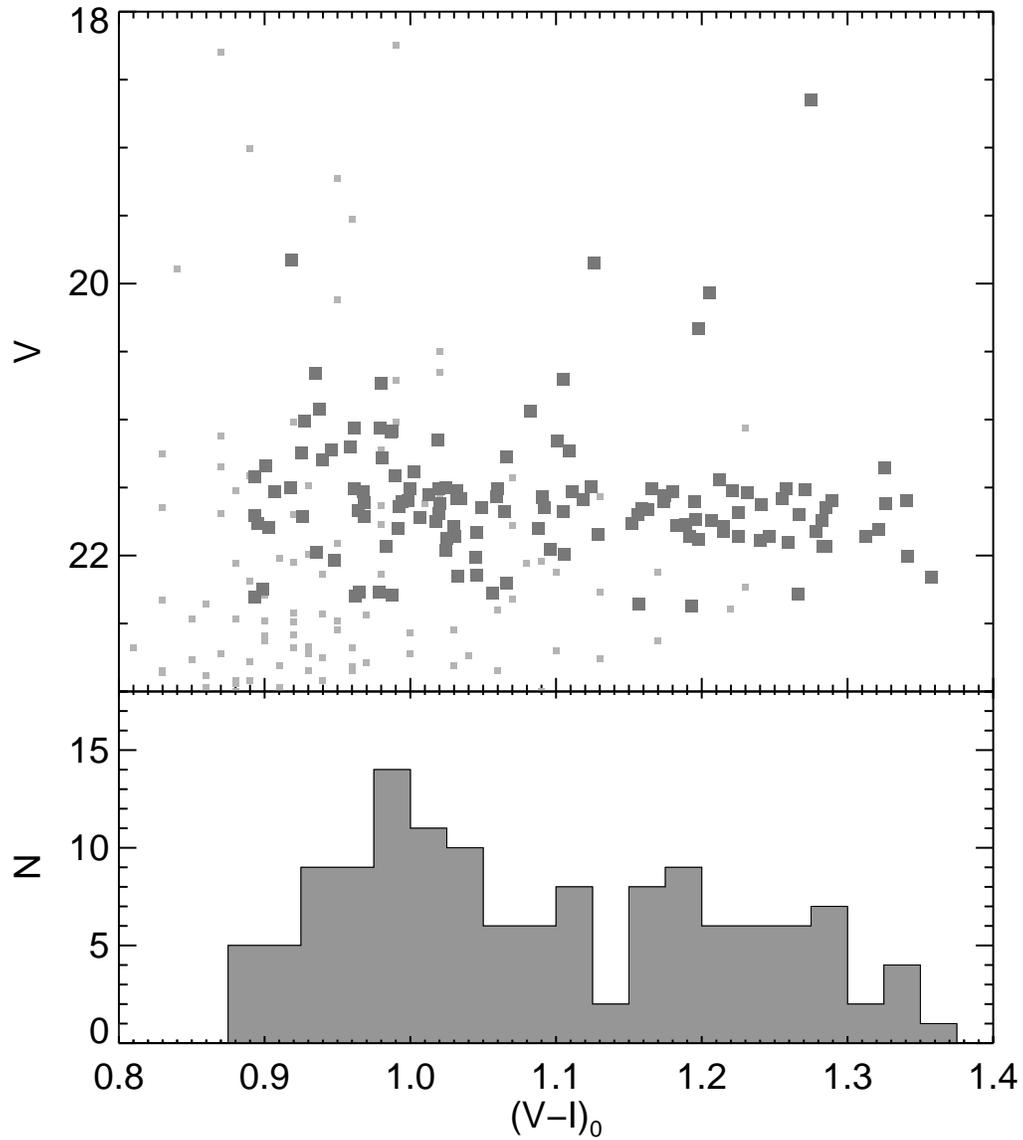}
\caption{The $V$ versus $V-I$ CMD (upper panel) and the color histogram (lower panel) show that our selected M87 GCs (dark gray filled circles) cover the typical color range of GCs 
(0.8$<\,V-I\,<$1.4). 
Subaru/Suprime-cam photometry \citep{tamura06a, tamura06b} are shown as light-gray filled circles.
\label{M87_1}}
\end{center}
\end{figure}

\begin{figure}[h!]
\begin{center}
\includegraphics[width=16.5cm]{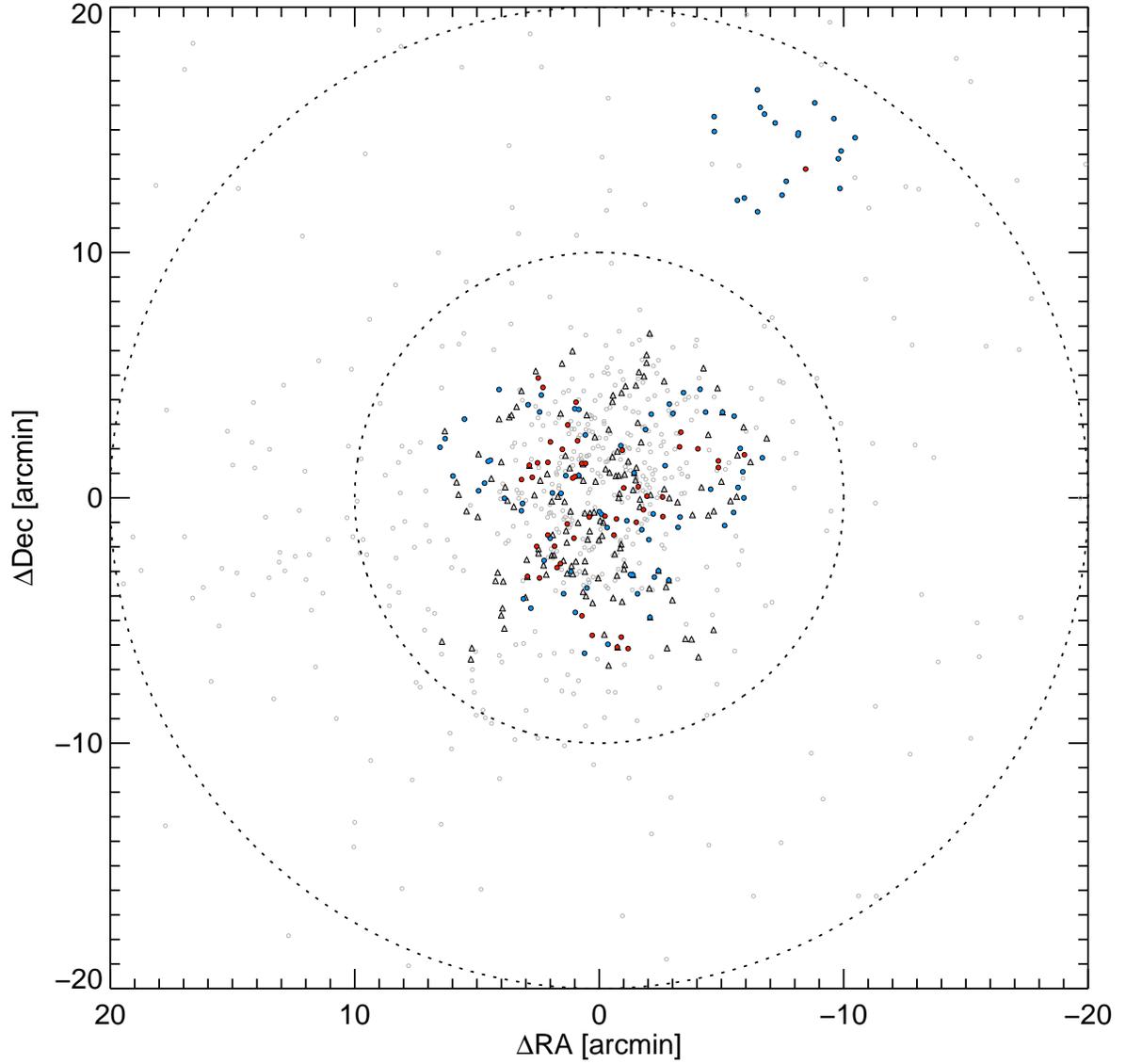}
\caption{The locations of the GCs are denoted as blue ($V-I\,<$\,1.1) and red ($V-I\,>$\,1.1) filled circles.
The dotted circles represent the distances 10$'$, and 20$'$ from the galactic center.
About 18$\%$ of the cluster targets are situated outside 10$'$. The gray circles denote the GC sample used in the kinematics study of the M87 GC system by \citet{strader11}. The triangles denote the positions of GCs in C98.
\label{M87_2}}
\end{center}
\end{figure}

The spectroscopic GC targets are selected from the photometric data on the M87 GC candidates obtained with Suprime-Cam on the 8.2 m Subaru telescope \citep{tamura06a, tamura06b}.
In Figure \ref{M87_1}, the $V$ versus $V-I$ color$-$magnitude diagram (CMD; upper panel) and the color histogram (lower panel) show that the selected clusters cover the typical color range of GCs (0.8 $<\,V-I\,<$1.4) and exhibit a clear color bimodality.
Figure \ref{M87_2} shows the locations of the GCs as blue ($V-I\,<$\,1.1) and red ($V-I\,>$\,1.1) filled circles.
In the inner target fields ($<$\,10$'$) where GC density is very high, we prioritized the red clusters with $V = 21.5-22.0$ mag because, at larger distances, color bimodality is less clear due to the decreasing contribution of red GCs. 
The magnitude range is chosen in order to avoid the possible effect of the known blue tilt of the M87 GC system at $V\le$\,21.5 \citep{strader06}.
The phenomenon of a lack of bright blue clusters is commonly observed in massive elliptical galaxies \citep{ostrov93, dirsch03, harris06, mieske06, strader06,  harris09,  mieske10}.

We performed spectroscopy on a total of 157 M87 GC candidates with the Multi-Object Spectrography (MOS) mode of the Faint Object Camera and Spectrograph (FOCAS) on the Subaru telescope on 2007 March $16-19$.
We used a configuration of a 300B grism and L600 filter resulting in a dispersion of $1.35$ {\small \AA}/pixel and a spectral coverage of $3700-6000$ {\small \AA}. 
Object spectra were obtained through $0.''5$ slits, rendering an overall spectral resolution of R $\sim\,800$ or 5.4 {\small \AA}, which is calculated using the spectral dispersion ($1.35$ {\small \AA}/pixel) and the FWHM of a sky line 4.0 pixel. 
Seeing was on average 1$''$, ranging between $0.''9$ and $1.''5$ over the four nights.
FOCAS has a circular 6$'$ field of view and the four masks contained between 28 and 46 objects (Table \ref{tbl1}).
The typical exposure time for the M87 fields was 1800 s, which adds up to the total integration time of $\sim\,$22 hr (Table \ref{tbl1}).
In addition, we observed a set of calibration stars including two flux standard stars (Feige34 and HZ44), three RV standard stars (HD68874, HD69148, and HD160952), and 26 Lick standard stars with a long-slit mode.
Dome flats and arc frames for both long-slit and MOS observations were taken in between exposures for object spectra.

We used basic spectroscopic data reduction procedures (bias subtraction, cosmic-ray removal, flat-fielding, field distortion correction, flexure effect correction,  wavelength calibration,  sky subtraction, spectrum extraction, extinction correction, and flux calibration) using the FOCASRED package.
We applied 3 pixel on-chip binning in the spatial direction and normalized the spectra by the median flux density at $4900-5100$ {\small \AA}. 
Finally, to further explore the CMRs of M87 GCs, we have also cross-identified our spectroscopic targets with the $M-T2$ Washington photometry data.
We obtained $M$ and $T2$ images of M87 GCs with the Mosaic II CCD imager of the 4 m Blanco telescope at the Cerro Tololo Inter-American Observatory (CTIO).
The Mosaic II CCD consists of eight 2048 $\times$ 4096 array with a pixel scale of 0.27, yielding  a field of view of 36\arcmin $\times$ 36\arcmin.
Images were preprocessed through IRAF\footnote{IRAF is distributed by the National Optical Astronomy Observatory, which is operated by the Association of Universities for Research in Astronomy, Inc., under cooperative agreement with the National Science Foundation.} using the MSCRED package \citep{valdes1998}.
The standard data reduction was then performed, and the photometry of the objects was obtained using the DAOPHOT II/ALLSTAR package \citep{stetson1987}.
We used standard stars by \citet{landolt1992} and \citet{stetson2000} for the photometric calibration.
Eight standard fields, each containing 15–25 standard stars with a wide color range, were taken at various airmasses.
Reddening correction and astrometry were done using the reddening maps by \citep{schlegel1998} and the USNO-B1.0 catalog \citep{monet2003}, respectively.

\section{Radial Velocity Measurement}

Among our selected sample cluster candidates, we now identify bona fide cluster members through the measurement of RVs of the candidate GCs.
RVs of the 157 GC candidates are measured by the Fourier cross-correlation technique \citep{tonry79}, using the FXCOR task in the IRAF RV package.
We fit the continuum of the spectra using the spline fit with $3\sigma$ clipping, then subtract the fit from the original spectra of GC candidates.
For the reference spectra required for this method, we observed three velocity standard stars of spectral types G8III during the observing run.
Then the continuum-subtracted spectra of GC candidates are cross-correlated with those of RV stars.
Each velocity standard star is separately used, yielding three independent values of RVs for GCs.
The derived velocity values using three different standard stars are consistent with each other within $1\sigma$,  and much smaller than the individual measurement errors.
Because the cross-correlation technique is rather sensitive to the noisy part of an input spectrum, which can smear the correlation signal, the choice of wavelength range to be cross-correlated is important.
We choose the wavelength region of $4000-5500$ {\small \AA} for the task in order to avoid the relatively low signal-to-noise-ratio (S/N) region of our object spectra below $\sim$\,4000 {\small \AA}, and a strong sky emission-line region around 5600 {\small \AA}.
This range includes $15-18$ Lick features.
In obtaining the final velocity values for GCs, the cross-correlated range of wavelengths was adjusted on an individual object basis.
Given the consistency between the velocity values derived with different velocity standard stars, the final RVs of GCs are measured using the RV standard star HD 68874 based on the quality of the standard star data.
The uncertainties of the RVs vary with the mask fields due to the S/N variation of the fields.
The uncertainties range mostly from $35-73$ km\,s$^{-1}$, with the typical error of $\sim$\,61 km\,s$^{-1}$ within 10$'$ fields.

The typical velocities of M87 GCs are known to be $200-2550$ km\,s$^{-1}$ based on the previous kinematic studies of the system \cite[C98;][]{hanes01}. 
Because objects with RV $<$ 200 km\,s$^{-1}$ are almost definitely foreground Galactic stars, we consider the velocity criterion of 200 $<$ RV $<$ 2550  km\,s$^{-1}$ for the GCs associated with M87.
Except for obvious contamination (emission-line galaxies and/or AGN) and objects whose spectra are useless due to bad columns, we find that 130 GCs satisfy this criterion.
The measured velocities of all cluster candidates are presented in  Table \ref{tbl2} along with the corresponding magnitude and color information by \citet{tamura06a, tamura06b} in the fourth and fifth columns.

\begin{figure}[h!]
\begin{center}
\includegraphics[width=15.5cm]{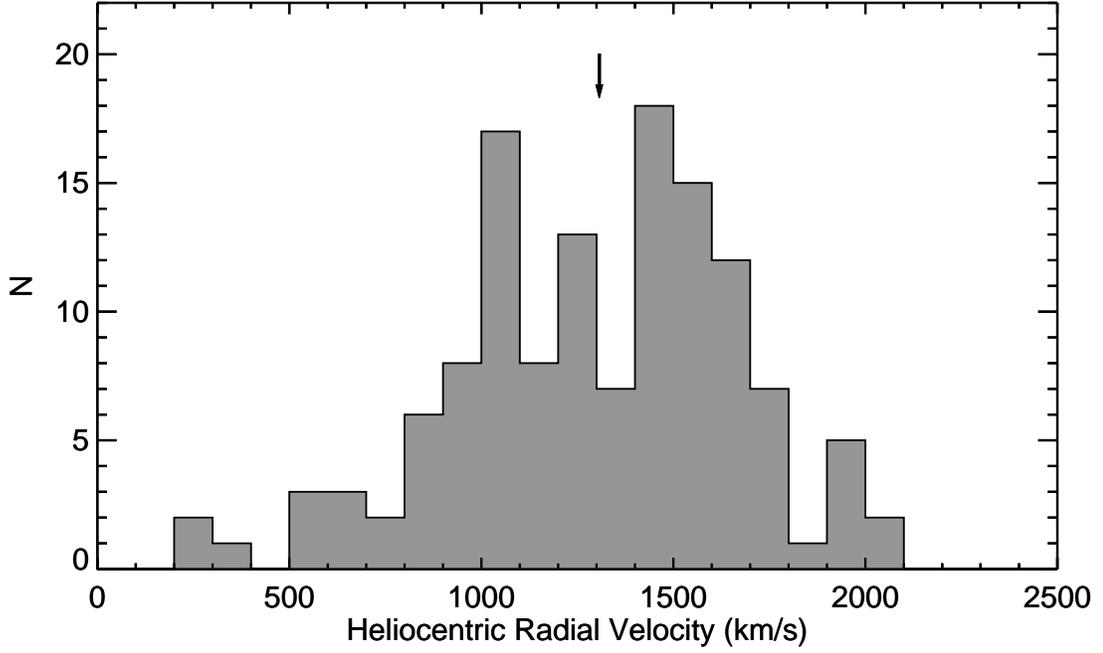}
\caption{Radial velocity histogram of the 130 confirmed M87 GCs. The value of the known radial velocity of M87 from the previous studies \cite[C98; S11;][]{hanes01} is 1307 km\,s$^{-1}$ at a distance of 16.1 Mpc, and is denoted by an arrow.}
\label{M87_3}
\end{center}
\end{figure}

The histogram of RVs of the confirmed GCs is shown in Figure \ref{M87_3}.
The RV distribution of M87 GCs has double peaks, and the clusters are distributed roughly symmetrically around the mean velocity of M87 GCs, 1299 km\,s$^{-1}$. This value is in good agreement with the known RV of M87 \cite[1307 km\,s$^{-1}$, C98;][hereafter S11]{hanes01, strader11} at a distance of 16.1 Mpc.

\begin{figure}[h!]
\begin{center}
\includegraphics[width=16cm]{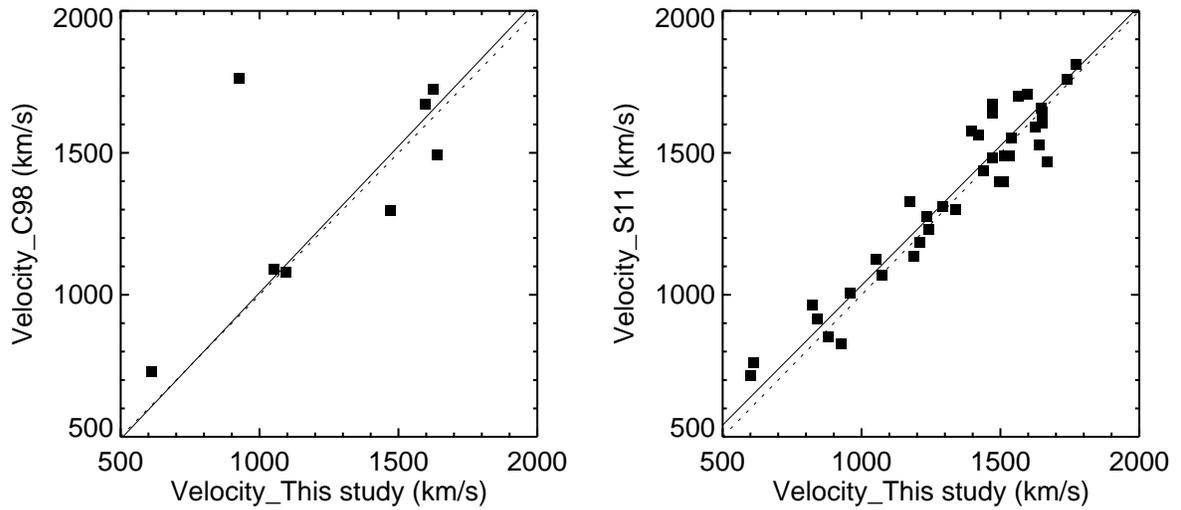}
\caption{The left panel shows eight GCs in common in the C98 catalog, and the right panel shows 39 GCs in common with new velocity measurements by S11. 
The solid and dotted lines represent the orthogonal least-squares fits to the data and the one-to-one relations, respectively. 
\label{M87_4}}
\end{center}
\end{figure}

Some of the GCs have previous measurements by C98 and S11.
Figure \ref{M87_4} illustrates these GCs in comparisons between each set of measurements. 
The left panel shows nine GCs in common with the C98 catalog, and the right panel shows 39 GCs in common with the new velocity measurements by S11.
The velocities of these common GCs are in general consistent.
The dotted line in each panel represents the least-squares fit to the data, and the rms of both fits are found to be comparable to the measurement errors.
One exception is F233 (RV = 962 km\,s$^{-1}$) which has measurements in all three catalogs.
The velocity value for this cluster by C98 is 1762 km\,s$^{-1}$.
The cause of this discrepancy is unknown.
Our value for this cluster is in good agreement with the more recent measurement by S11.

\section{Lick Index Measurement}

The Lick/IDS system of absorption-line indices provides a reference frame in which one can derive metallicities, ages, and abundance ratios for stellar populations. 
The Lick/IDS system was originally created by \citet{burstein84}, and has been continuously revised and improved over the years \cite[W94;][]{trager98}.
The system has also been tried and developed in SSP modeling, with their variations with stellar age, metallicity, and $\alpha$-element abundance considered \cite[e.g.,][]{bruzual03, thomas03, chung13}.
The observed Lick index values are compared to the SSP Lick indices to estimate ages, metallicities, and [$\alpha$/Fe] ratios.

\begin{figure}[h!]
\begin{center}
\includegraphics[width=14.5cm]{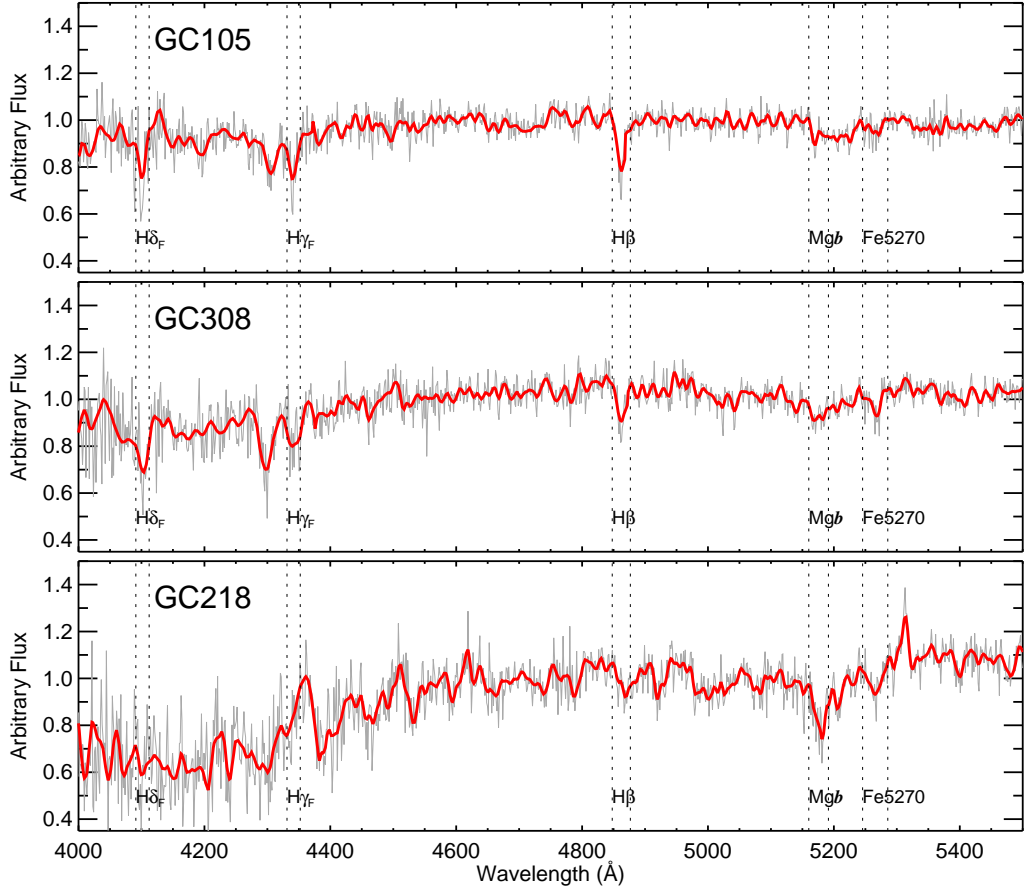}
\caption{Representative rest-frame spectra with RV = 0 km\,s$^{-1}$.} 
The vertical dotted lines mark the locations of the absorption features, the names of which are denoted. 
The metallicity [Fe/H] of the spectra increases from the top to  bottom panel. 
The smoothed spectra that match the original Lick resolutions (W97) are shown in red.
\label{M87_5}
\end{center}
\end{figure}

Figure \ref{M87_5} shows three representative spectra of GCs with RV = 0 km\,s$^{-1}$ and the marked locations of the absorption features (gray solid lines).
The metallicity [Fe/H] of the spectra increases from the top to bottom panel.
Following the  wavelength-dependent Lick spectral resolution, we smoothed our spectra (red solid lines) to match the original Lick resolutions (W97).
All indices and errors were measured using the C$^{++}$ program Indexf \citep{cardiel98}.
In this study, we adopt the passband and pseudo-continuum definitions by W94 and W97.
The object spectra and error spectra were provided as inputs for Indexf Lick index measurements.
Indexf generates Lick index uncertainties by performing 100 Monte Carlo simulations; indices are measured on the Poisson noise added spectra, and the 1$\sigma$ standard deviation of the measured index is taken as the Lick index uncertainty.

\begin{figure}[h!]
\begin{center}
\includegraphics[width=16.5cm]{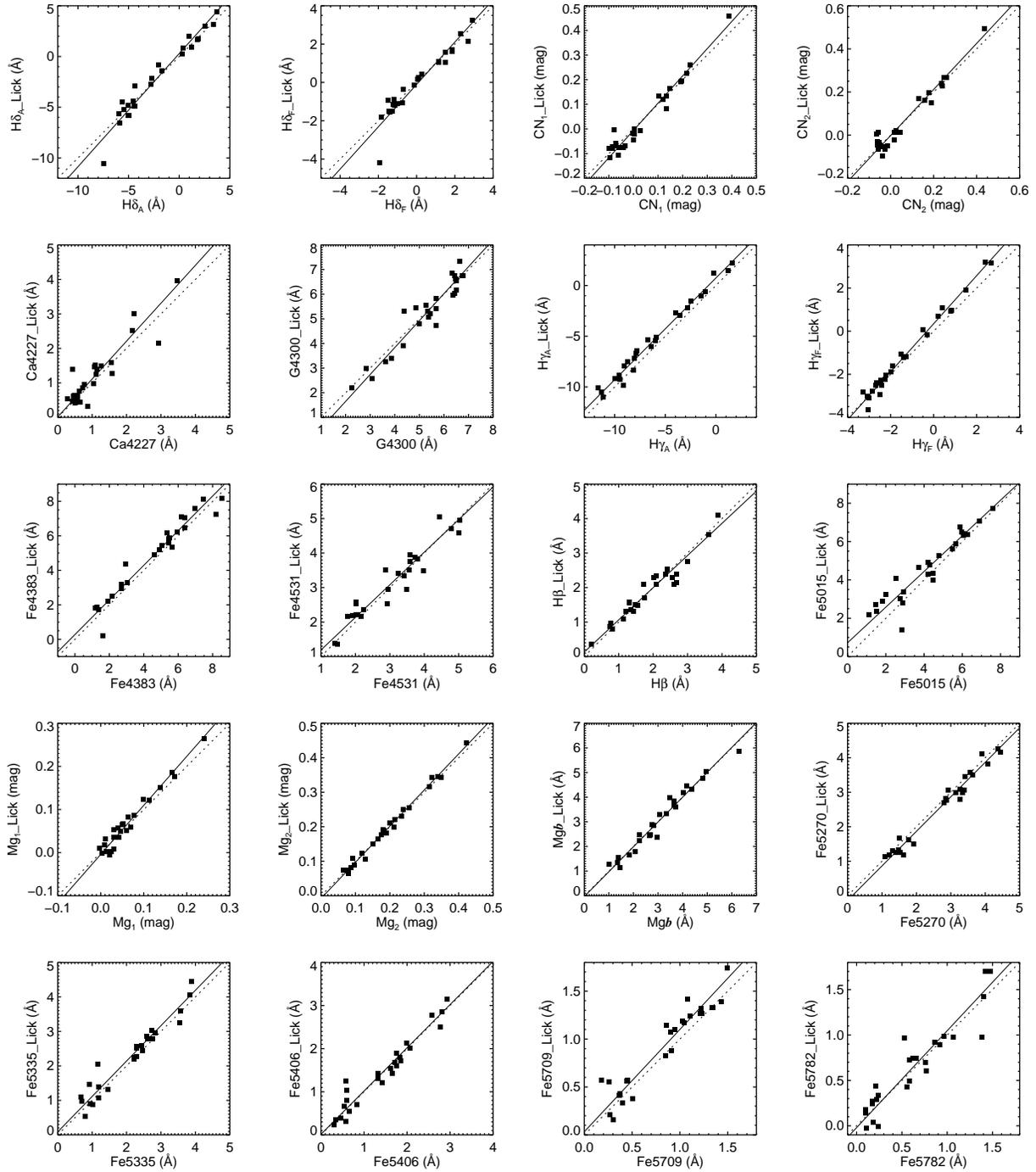}
\caption{The comparison between the Lick indices of 26 Lick/IDS standard stars in the Lick system (W94, W97) and our measurements. 
The solid and dotted lines represent the orthogonal least-squares fits and the one-to-one relations, respectively.
\label{M87_6}}
\end{center}
\end{figure}

In Figure \ref{M87_6}, we calibrate our spectra to the Lick system by comparing our measured indices for the 26 Lick standard stars to the Lick/IDS measurements.
The offsets between the two measurements were then applied to the Lick indices of our 130 M87 GCs listed in Table \ref{tbl2}.
The measured Lick indices and their uncertainties are given in Tables \ref{tbl3} and \ref{tbl4}, respectively. In addition to the above Lick indices,
we have also combined individual indices to create a more robust composite index, such as
\begin{equation}
{\mathrm{[MgFe]' = \sqrt{Mgb~[0.72 \times Fe5270 + 0.28 \times Fe5335]}}}
\label{eqn1}
\end{equation}
\citep{thomas03} and
\begin{equation}
{\langle\mathrm{Fe}\rangle = \mathrm{(Fe5270 + Fe5335) / 2}}.
\label{eqn2}
\end{equation}

\begin{figure}[h!]
\begin{center}
\includegraphics[width=16cm]{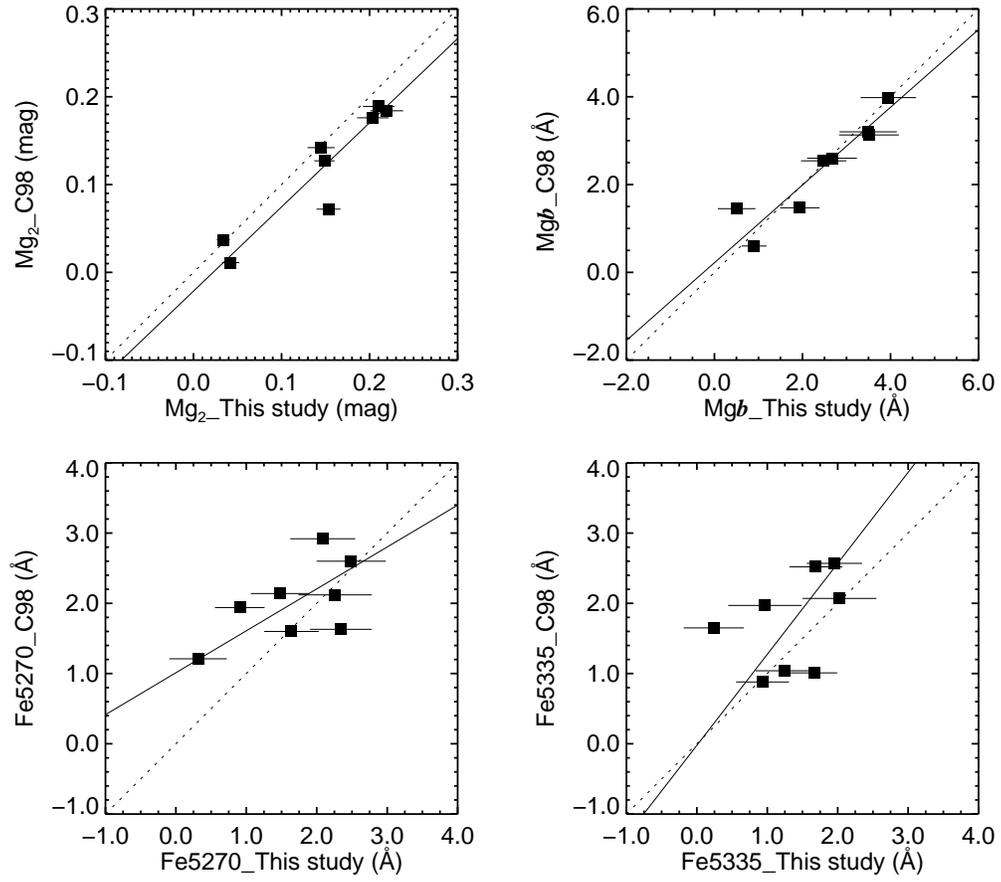}
\caption{The comparison between the Lick index measurements from the common GCs by C98. 
The solid and dotted lines represent the orthogonal least-squares fits and the one-to-one relations, respectively.
\label{M87_7}}
\end{center}
\end{figure}

We show the comparisons of our Lick index measurements with those by C98 in Figure \ref{M87_7}. 
Among the 150 GCs obtained by C98, 10 GCs are also on our slit masks and the Lick indices were successfully measured for 9 of these.
After excluding one GC with inconsistent RV values, Lick indices of eight common GCs are compared.
While Mg$b$ shows good agreement between the two data sets, the discrepancies are apparent for Fe and Balmer indices. 
We note that the Lick indices by C98 are measured using the older definitions by \citet{burstein84}, \citet{faber85} and \citet{gorgas1993}, while ours adopts the definitions by \citet{trager98}. This may be partly responsible for some of the larger rms values.

\section{Age, Metallicity, and Elemental Abundance}

In this section, we explore the stellar population properties---ages, metallicities, and elemental abundance ratios---for our sample GCs via various methods and check for consistency between the results.
In Section 5.1, we first examine the general ranges and trends of GC age, metallicity, and abundances in the index$-$index diagnostic diagrams by comparing the observations to the SSP model grids.
Then, in Section 5.2, we  obtain the purely empirical metallicity values, which are independent of models, using the most recent comprehensive catalog of Milky Way globular clusters \cite[MWGCs,][]{kim16}. These MWGCs are old and of solar abundance ratios.
Next, in Section 5.3, in order to assign ages, metallicities, and elemental abundance ratio values to each individual GC we adopt the method of multi-index fitting to a set of SSP models via $\chi^2$ minimization described by \citet{proctor04}. 
The derived values in Section 5.3 are used in the analysis for the rest of the paper.

\subsection{Absorption-line Diagnostic Plots} 

The index$-$index diagnostics provide a simple, qualitative picture on the distributions of ages, metallicities, and elemental abundance ratios of a GC system.
In Figure \ref{M87_8}, we present plots with selected indices as a function of the metallicity-sensitive composite index [MgFe]$'$ (Equations \ref{eqn1}).
The combination of indices containing the three Balmer indices H$\beta$, H$\gamma_{A}$, and H$\gamma_{F}$ represent the age$-$metallicity distributions of the clusters.
The superimposed grids are the model predictions from the Yonsei Evolutionary Population Synthesis (YEPS) model \citep{chung13} for ages 2, 3, 5, 8, 12, and 15 Gyr, and for metallicities [Fe/H] = $-2.5, -1.5, -0.5, 0.0,$ and 0.5\footnote{The full set of the spectrophotometric model predictions including model absorption indices are available online at http://cosmic.yonsei.ac.kr/YEPS.htm.}.
The black (gray) grids correspond to the models with [$\alpha$/Fe] = 0.3 ([$\alpha$/Fe] = 0.0).
The fiducial YEPS model assumes the \citet{Salpeter1955} initial mass function of $s=2.35$.
The M87 GCs are shown as filled gray circles and the 53 MWGCs by \citet{kim16} as open squares.
M87 GC system exhibits a wide range of metallicities, mostly placed between [Fe/H] = $-2.5$ and $0.0$.
While the majority of GCs fall on the old age lines, 
some ($5-10$\,$\%$) clusters appear young depending on the choice of Balmer line, as they are placed on the $3-5$ Gyr model lines.

To estimate the abundance ratios [$\alpha$/Fe] of the clusters, we choose tracers of $\alpha$-elements (Mg{\it b}) and iron-peaked elements ($\langle$Fe$\rangle$, Equations \ref{eqn2}) (lower-right panel). 
$\alpha$-elements and more than two-thirds of iron-peak elements are produced by Type II and Type Ia supernovae (SNe), respectively.
By virtue of the distinct characteristics of each SN type in terms of timing and duration of events, the ratio of [$\alpha$/Fe] provides clues to the star-formation history of the host.
Enhanced [$\alpha$/Fe] ($\sim$\,0.3) is commonly found for GCs in early-type galaxies \citep{worthey92,matteucci94,trager00,thomas05}, indicating rapid star formation.
The observational points are denoted with the symbols as before, and the 12 Gyr model GCs with varying [$\alpha$/Fe] ratios of 0.0 (dotted line), 0.3 (solid line), and 0.6 (dashed line) are overlaid.
The majority of M87 GCs lie within the range of [$\alpha$/Fe] = $0.0-0.3$ which is consistent with the previous findings on the extragalactic GC system for early-type galaxies. At the low-metallicity end, it is difficult to draw a conclusion on the distribution of [$\alpha$/Fe] due to convergence of the model grids as well as somewhat large observational uncertainties.

We next examine the abundance pattern of carbon and nitrogen through a set of index$-$index diagnostics of M87 GCs using CN$_{2}$ and G4300 indices (Figure \ref{M87_9}).
In the left panel, the CN$_{2}$ distribution of M87 GCs are shown as light-gray filled squares, and the clusters with higher S/N are plotted as dark gray squares.
Although there is a large scatter of M87 data compared to the MWGCs \cite[open squares;][]{kim16}, a general trend with metallicity is obvious.
The overplotted model grids represent the varying [$\alpha$/Fe]  ratios of solar (gray) and 0.3 dex (black), respectively.
CN$_{2}$ exhibits systematically stronger index strengths for both MWGCs and M87 GCs across the full metallicity range with respect to the model predictions.
This indicates carbon and/or nitrogen overabundances for the two GC systems, the trend of which seems to be common in typical GC systems \citep{puzia05, cenarro07, beasley08, woodley10}.
In the right panel, G4300 is known to be mainly sensitive to carbon \citep{tripicco95} and shows a smaller offset between model predictions and the observation.

The CN-strong features observed in GCs are ascribed to several possible sources, including pollution of gas by intermediate-mass AGB stars \citep{cottrell81, dantona07}, early enrichment achieved by the stellar wind of fast-rotating massive stars \citep{maeder06, prantzos06, decressin2007}, and difference in surface gravity \citep{mucciarelli2015, lee2016, gerber2018}, opacity \citep{macLean2018} and stellar mass function \citep{renzini2008, kim2018}.
The nitrogen overabundance may also be attributed to the two generations of the constituent stars of GCs \citep{carretta08, dercole08}.
In this scenario, second-generation stars are formed from the material pre-enriched by supernovae explosions.
Because the observed G4300 indices with higher S/N do not seem to stray far from the models, we can assume that carbon enhancement is not likely. 
Hence, overenrichment of nitrogen in M87 GCs may be a source of the significant discrepancy between the observation and model.
Due to large uncertainties, we cannot conclusively determine the abundance pattern of M87 GCs, but the overall trends of CN$_{2}$ and G4300 against the metal-sensitive index [MgFe]$'$ indicate an overabundance in carbon and nitrogen.

\subsection{Empirical Metallicities} 

We employ a simple, model-independent approach to assigning metallicities to individual M87 GCs in which MWGCs are used as metallicity calibrators.
It is noted in \citet{beasley08} and \citet{strader2004} that, although mainly tied to the \citet{zinn1984} metallicity scale, there is no metallicity scale universally agreed upon.
In this study, we refer to our empirically derived metallicity as [M/H] which may not strictly reflect either [Fe/H] or overall metallicity ([Z/H] (see \citet{beasley08} for a more detailed discussion).
We use spectroscopic data on 53 MWGCs \citep{kim16} which are calibrated onto the Lick system and thus directly comparable to our M87 GC data. 

\begin{figure}[h!]
\begin{center}
\includegraphics[width=16cm]{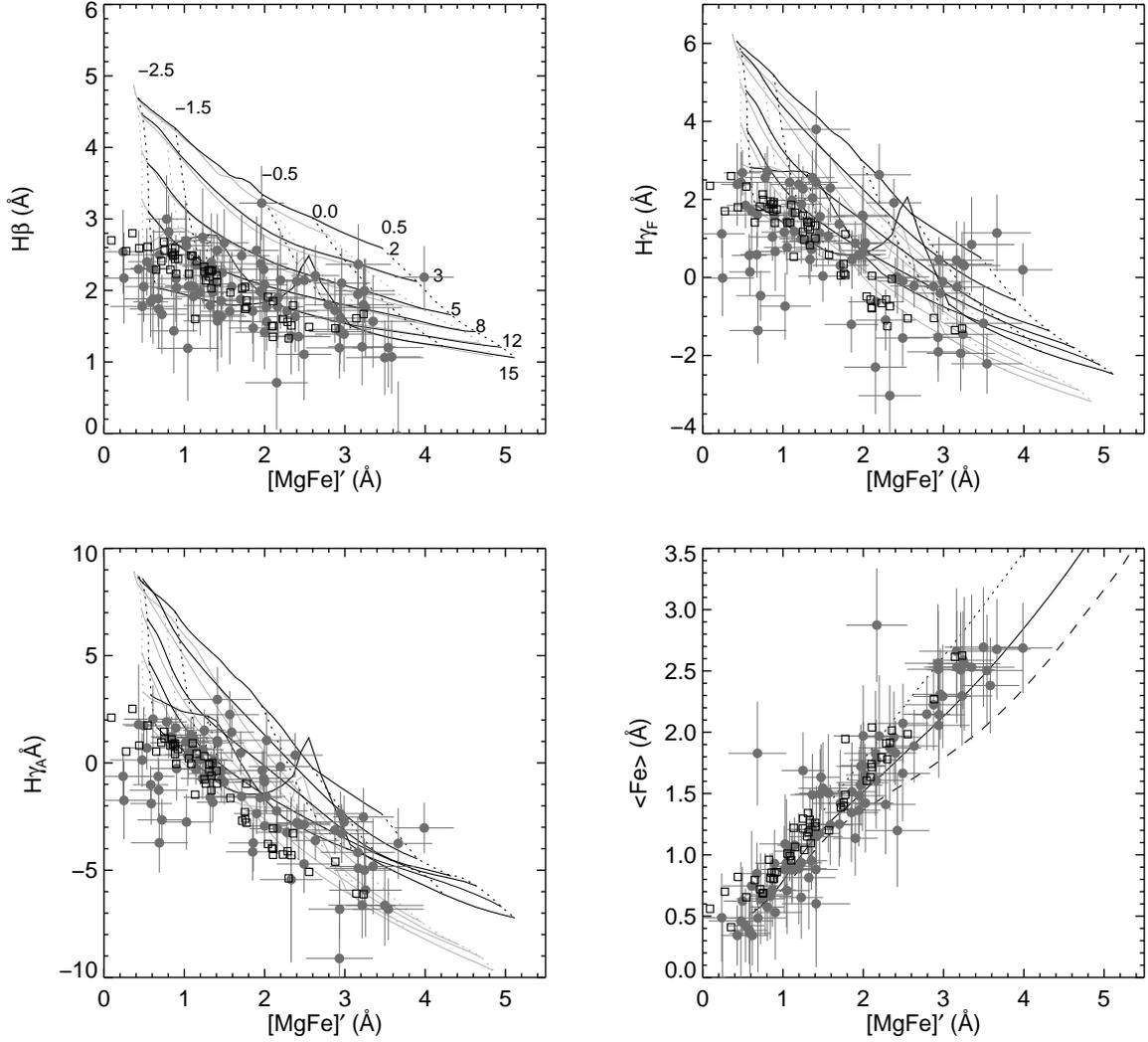}
\caption{Index$-$index diagnostics for M87 GCs and MWGCs. 
The indices of the two GC systems are compared to the YEPS model grids \citep{chung13} with [$\alpha$/Fe] = 0.3 (black lines) and [$\alpha$/Fe] = 0.0 (gray lines). 
The isoage lines correspond to 2, 3, 5, 8, 12, and 15 Gyr, and the isometallicity lines are of [Fe/H] = $-2.5, -1.5, -0.5, 0.0,$ and 0.5 in the Balmer$-$[MgFe]$^\prime$ planes. 
The model lines in the $\langle$Fe$\rangle$$-$[MgFe]$^\prime$ plane represent the 13.5 Gyr old GCs with [$\alpha$/Fe] = 0.0 (dotted line), 0.3 (solid line), and 0.6 (dashed line). 
M87 GCs are shown as filled gray circles and MWGCs by \citet{kim16} are shown as open squares.
\label{M87_8}}
\end{center}
\end{figure}

\begin{figure}[h!]
\begin{center}
\includegraphics[width=16cm]{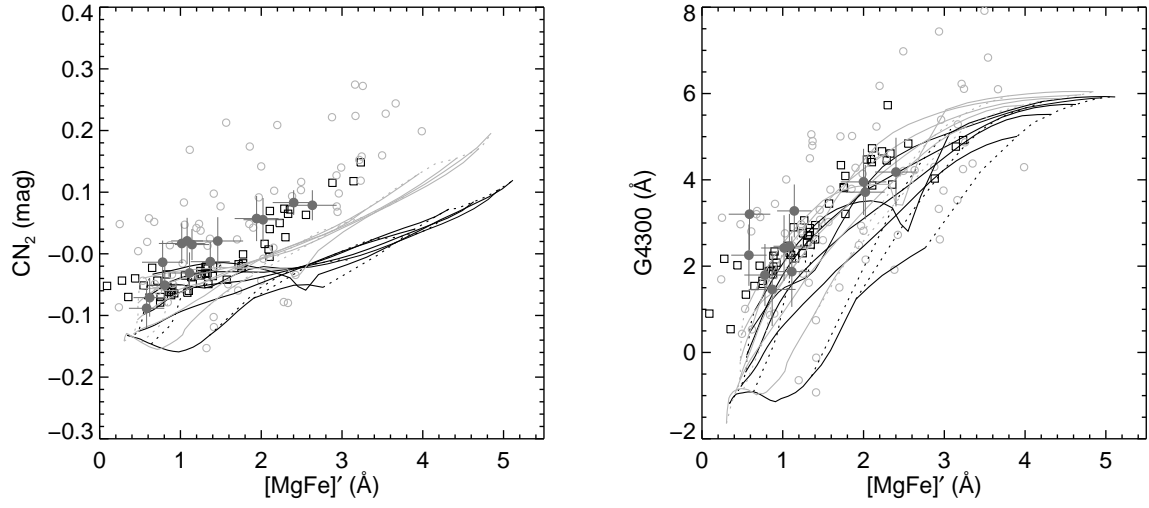}
\caption{Index$-$index diagnostics for M87 GCs and MWGCs. 
Same as Figure \ref{M87_8}, but with CN$_{2}$ and G4300. 
M87 GCs with higher S/N ($\,>\,10$) are shown as filled gray circles.
Open gray circles are the rest of the M87 GCs.
\label{M87_9}}
\end{center}
\end{figure}

\begin{figure}[h!]
\begin{center}
\includegraphics[width=15.5cm]{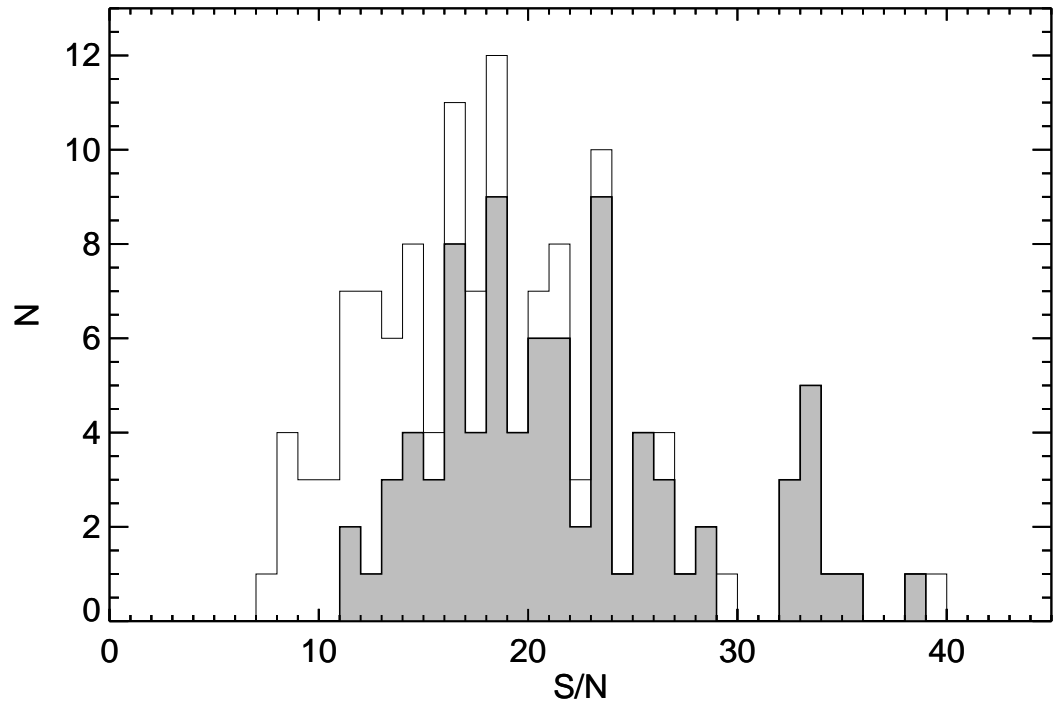}
\caption{S/N distributions of 130 confirmed M87 GCs (open) and 83 old GCs with higher S/N (\,$>$\,10, gray) used in the abundance analysis.}
\label{M87_11}
\end{center}
\end{figure}

We adopt the method that closely follows that of \citet{beasley08}. 
In order to derive empirical metallicities, we obtain correlations between metallicities provided by the Harris catalog \citep{harris1996} and the measured Lick indices of 53 MWGCs \citep{kim16}.
To better define the relationships that exhibit a hint of nonlinearity, we fit the multiorder polynomial to the data using orthogonal distance regression \citep[ODR;][]{jefferys88}.
In most cases, the correlations are characterized by second-order polynomial, except for H$\beta$ and CN indices, which are better defined by higher-order correlations. 
Figure \ref{M87_10} illustrates the resulting fits for the 18 Lick indices and the fitted coefficients are listed in Table \ref{tbl5}.
Nine indices are chosen for the derivation of empirical metallicities based on both the goodness of the ODR fits on the MWGCs and the index measurement qualities for M87 GCs. 
These indices include Fe4383, Fe4531, H$\beta$, Fe5015, Mg{\it b}, Mg$_{2}$, Fe5270, Fe5335, and Fe5406.
As shown previously in Figure \ref{M87_8}, young GCs cannot be readily separated due to the possible effect of hot horizontal-branch stars present in the GCs.
Taking this into consideration, to examine the properties of the old members of the M87 GC system exclusively, we leave out four clusters that are found to be $\,\le\,8$ Gyr in all three Balmer$-$[MgFe]$^\prime$ planes (Figure \ref{M87_8}).
As the spectroscopic ages of M87 GCs and stellar population are in general old (\,$>$\,12 Gyr, \citet{cohen98, kuntschner2010}), we assume the rest are old GCs. 
For 83 old GCs with sufficient quality (10\,$<$\,S/N\,$<$\,40) for an abundance analysis (Figure \ref{M87_11}), we obtain empirical metallicities [M/H] using the fitted coefficients for the chosen indices and list the values in Table  \ref{tbl3}.

\subsection{Multi-index Fits via \texorpdfstring{$\chi^2$}{chi2} Minimization}

For a more quantitative analysis on the metallicities, ages, and [$\alpha$/Fe] of M87 GCs, we perform $\chi^2$ multi-index fitting with the YEPS SSP models \citep{chung13}.
This technique, introduced and explained in detail by \citet{proctor04}, simultaneously fits a set of Lick indices to the model grids of [Fe/H], ages, and [$\alpha$/Fe] until $\chi^2$ is minimized.
The method has the advantage in that by using multiple absorption-line indices, it is more robust against individual index uncertainties.
We adopt the same method of rejecting the deviant indices and recalculating the $\chi^2$ statistics as described in \citet{proctor04}. 
Note that we exclude several indices (e.g., CN$_{1}$, CN$_{2}$, Ca4227 and G4300) from the initial fitting process because of their inconsistency with respect to the models.
We determined 13.5 Gyr to be the best-fit model age parameter to represent the old clusters in M87 and thus fixed the model as such in our multi-index fitting process to obtain the metallicities [Fe/H] and abundance ratios [$\alpha$/Fe]. 
We perform a $\chi^2$ routine with the same 83 GCs as the previous section.
The [Fe/H] values are compared to the empirical metallicities derived in Section 5.2 (Figure \ref{M87_12}) and found to be in good agreement with each other. 
Because the empirical [M/H] is set by MWGCs, we might expect offsets for GCs with [$\alpha$/Fe] higher than MWGCs' abundance ratios (see the bottom-right panel of Figure \ref{M87_8}).
GCs with higher $\alpha$-element abundance ratios ([$\alpha$/Fe]$\,>\,0.5$) are denoted with gray squares and they exhibit larger inconsistencies in the metal-poor region.
These GCs are excluded from the least-squares fits to the data.
The derived values of ages, metallicities, and abundance ratios for 87 GCs in this section are used in the analysis for the rest of the paper.

\section{Tests for Nonlinear-CMR Scenario for GC Color Bimodality}

\subsection{Color\texorpdfstring{$-$}{-}Metallicity Relations}

Our high-quality, homogeneous dataset enables us to test the theoretical prediction of inflected CMRs.
Figure \ref{M87_13} presents the derived spectroscopic metallicities [Fe/H] as a function GC colors $V-I$ (left panel) from Subaru/Suprime-cam \citep{tamura06b} and $M-T2$ (right) from CTIO.
The Suprime-cam photometry gives information for all the 83 GCs because the spectroscopic candidates are selected from it. The CTIO data provide 78 matches with our spectroscopic [Fe/H].
The shades of symbols represent the size of observation errors such that darker squares are GCs with smaller uncertainties. 

This figure is similar to Figure \ref{M87_2} of Paper II where the model CMR was explored for a combined heterogeneous data set of GCs in M87, M49, and the MW, but it now uses the by far more homogeneous dataset of M87 GCs only.
We find that the CMRs of M87 GCs in both metallicity$-$($V-I$) and metallicity$-$($M-T2$) planes display wavy features where slopes of the relations change at around [Fe/H]\,=\,$-1$.
The observed significant inflection along the optical CMRs supports the notion that the nonlinear CMRs between intrinsic metallicity and its proxy, colors, holds the key to the color bimodality phenomenon.
The effect of this nonlinear conversion from metallicities to colors will be demonstrated in Section 6.3. 

\subsection{Metallicity Distribution Functions}     

Here we combine our metallicities derived with the SSP model-fitting method with C98 metallicities in order to have a better understanding of the intrinsic metallicity distribution function (MDF) shape of the M87 GC system.
The left panel of Figure \ref{M87_14} illustrates the transformation of C98 metallicities to our M87 metallicities. 
The C98 catalog comprises bright GCs, which occupy regions similar to as our sample within 10$'$ of the galactic radius (Figure \ref{M87_2}).
The solid line represents the orthogonal least-squares fit, and the derived transformation is given by 
\begin{equation}
{\mathrm{[Fe/H]}_\mathrm{This\,\,study}} = -0.0097 + 0.9418\,{\mathrm{[Fe/H]}_\mathrm{C98}}\,.
\end{equation}
In Tables \ref{tbl6}, we list the transformed [Fe/H] of C98.

In the right panel, the distribution of the M87 GC metallicities from this study is shown as a red hatched histogram.
The blue hatched histogram displays the resulting metallicity distribution of C98 GCs with high-S/N C98 shifted to the same scale as that of ours.
The gray filled histogram represents the combined metallicities of Subaru and Keck spectroscopy.
The combined list of M87 GCs contains 98 higher-S/N GCs from C98 and 87 GCs from our study.
For  the GCs present in both catalogs, we take the values from our study.
This is one of the largest spectroscopic data on M87 GCs and of quality that allows reliable abundance study on the system.
A Gaussian mixture model (GMM) test on the observed MDFs (Table \ref{tbl7}) gives a high probability of the distribution being a unimodal Gaussian one ($P(\chi^2)=0.246$).

\subsection{Color Distribution Functions}     

In Figure \ref{M87_15} we simulate the M87 GC colors and compare them with the observations. 
The top row presents the observations and the best-fit model predictions (13.5 Gyr and [$\alpha$/Fe] = 0.3) for the CMRs of M87 GCs. 
The models trace the features present in the observations well.
In order to demonstrate the nonlinear-CMR hypothesis at work, we carry out the conversion process from metallicities to colors via our theoretical metallicity$-$color relations.
For an MDF, we make a simple assumption of a single Gaussian distribution of $10^{6}$ model GCs (top row, along the y-axes) to avoid small number statistics.
The assumed mean [Fe/H] and $\sigma$([Fe/H]) of $-$0.85 and 0.5 are similar to those of observed distribution in Figure \ref{M87_13}.
In the second row, we plot the color$-$magnitude diagrams of 5000 randomly selected model GCs for the $V-I$ and $M-T2$ colors as an additional aid to visualize the simulated color distributions.
We take observational uncertainties into account in the simulations.
The inflection in our theoretical [Fe/H]$-$color relations has the effect of projecting the equidistant metallicity intervals onto wider color intervals, causing scarcity in the color domain near the quasi-inflection point on each [Fe/H]$-$color relation.
As a result, the division into two groups of model colors are visible, which agrees with the observations (Figure \ref{M87_1}).
The resultant color histograms (third row) show bimodal distributions with stronger blue peaks, and with the paucity at quasi-inflection points translated to the dip positions for both colors.
Therefore, the theoretical prediction and the observed distributions of GC colors (bottom row) are consistent with each other.
This result is also in good agreement with that of Paper II on the general behaviors of M87 GCs.

In order to quantify the visual impression of the distributions, we perform a GMM test on the simulated and observed color histograms. 
Table \ref{tbl7} presents the resulting GMM outputs.
Bimodality is preferred for color distributions of the observed and simulated GCs ($P(\chi^2) = 0.001$).
The fraction of GCs assigned to the blue group is slightly higher in simulations.
We add vertical dotted lines in Figure \ref{M87_15} at the blue and red peaks for the observed histograms (fourth row) to better guide the eye for a comparison with the simulation (third row). We attribute the slight offset in red peak locations for $M-T2$ between the observed and simulated color histograms to several contributing factors, such as different choices of model ingredients involved.
Hence, we put more weight on the consistent histogram morphology between the observations and simulations.

\section{Summary and Discussion} 

\begin{figure}[h!]
\begin{center}
\includegraphics[width=16cm]{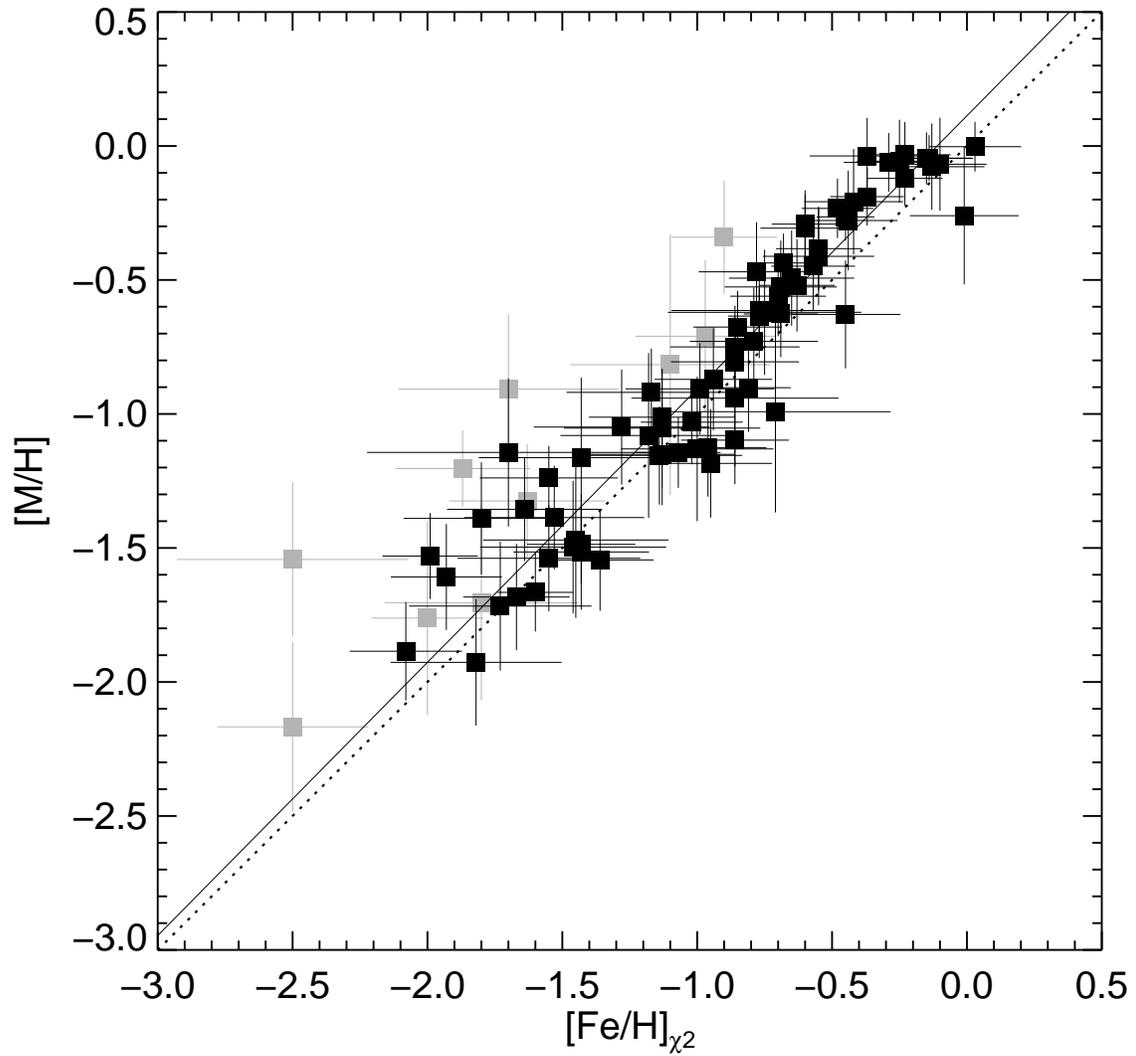}
\caption{Comparison between the M87 GC metallicities derived empirically \cite[y-axis;][]{beasley08} and by using SSP models \cite[x-axis;][]{proctor04}. 
The solid and dotted lines represent the orthogonal least-squares fits to the black squares and the one-to-one relations, respectively.
Gray squares are the GCs with higher alpha-elemental abundance ratios ([$\alpha$/Fe]\,$>$\,0.5.).}
\label{M87_12}
\end{center}
\end{figure}

We have examined the fundamental stellar population properties of M87 GCs and reported spectroscopic metallicity estimates using the homogeneous spectroscopic data.
We have adopted empirical and theoretical methods in estimating the individual GC metallicity values and found them consistent with each other (Figure \ref{M87_12}).
The empirical metallicity was obtained using the most recent comprehensive catalog of MWGCs \citep{kim16}.
The theoretical metallicity was based on multi-index fitting to a set of SSP models via $\chi^2$ minimization described by \citet{proctor04}. 

\begin{figure}[h!]
\begin{center}
\includegraphics[width=17cm]{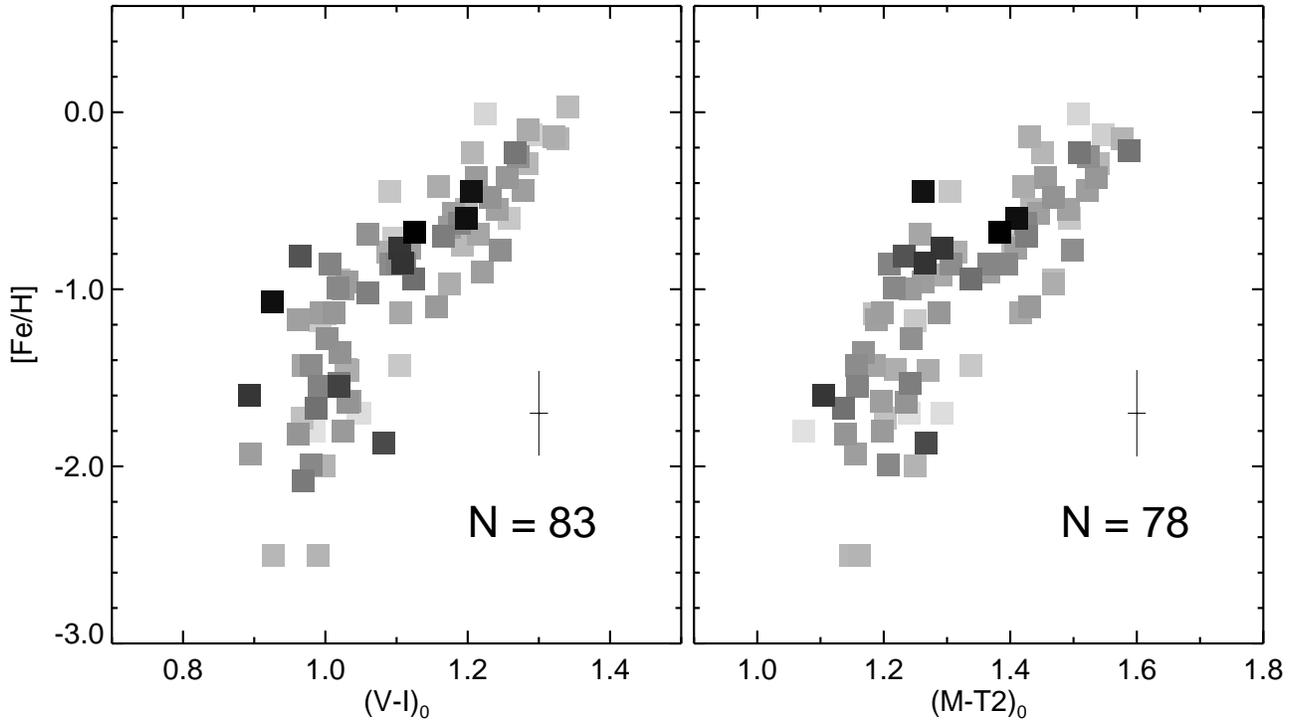}
\caption{The Subaru $V-I$ (left) and CTIO $M-T2$ (right) versus the spectroscopic metallicity relations of M87 GCs.
The shade of the squares corresponds to the observational errors. i.e., darker squares are GCs with smaller observation uncertainties.
The error bars show the typical observational errors in [Fe/H] and color. 
\label{M87_13}}
\end{center}
\end{figure}

\begin{figure}[h!]
\begin{center}
\includegraphics[width=18cm]{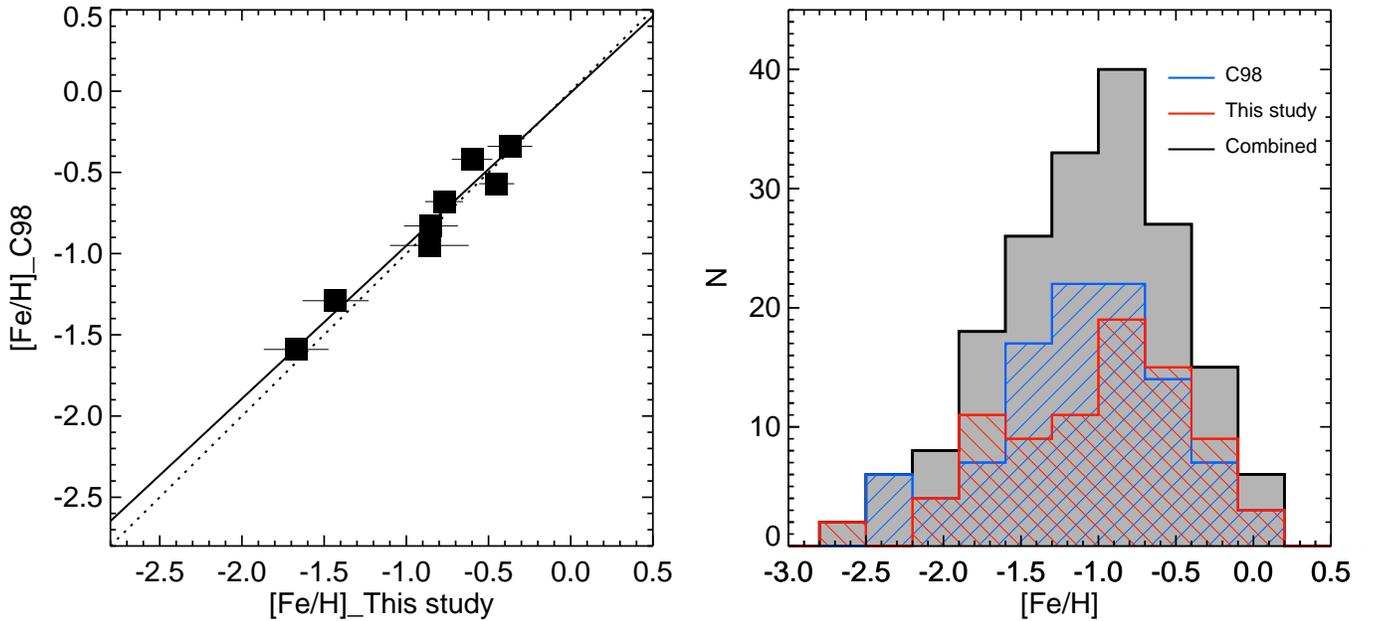}
\caption{(Left) Transformation of the C98 metallicity to our M87 metallicity. 
The solid and dotted lines represent the orthogonal least-squares fits and the one-to-one relations, respectively.
(Right) The red, blue, and gray histograms show the metallicity distributions of this study, C98, and the combined metallicities, respectively.
\label{M87_14}}
\end{center}
\end{figure}

\begin{figure}[h!]
\begin{center}
\includegraphics[width=13.5cm]{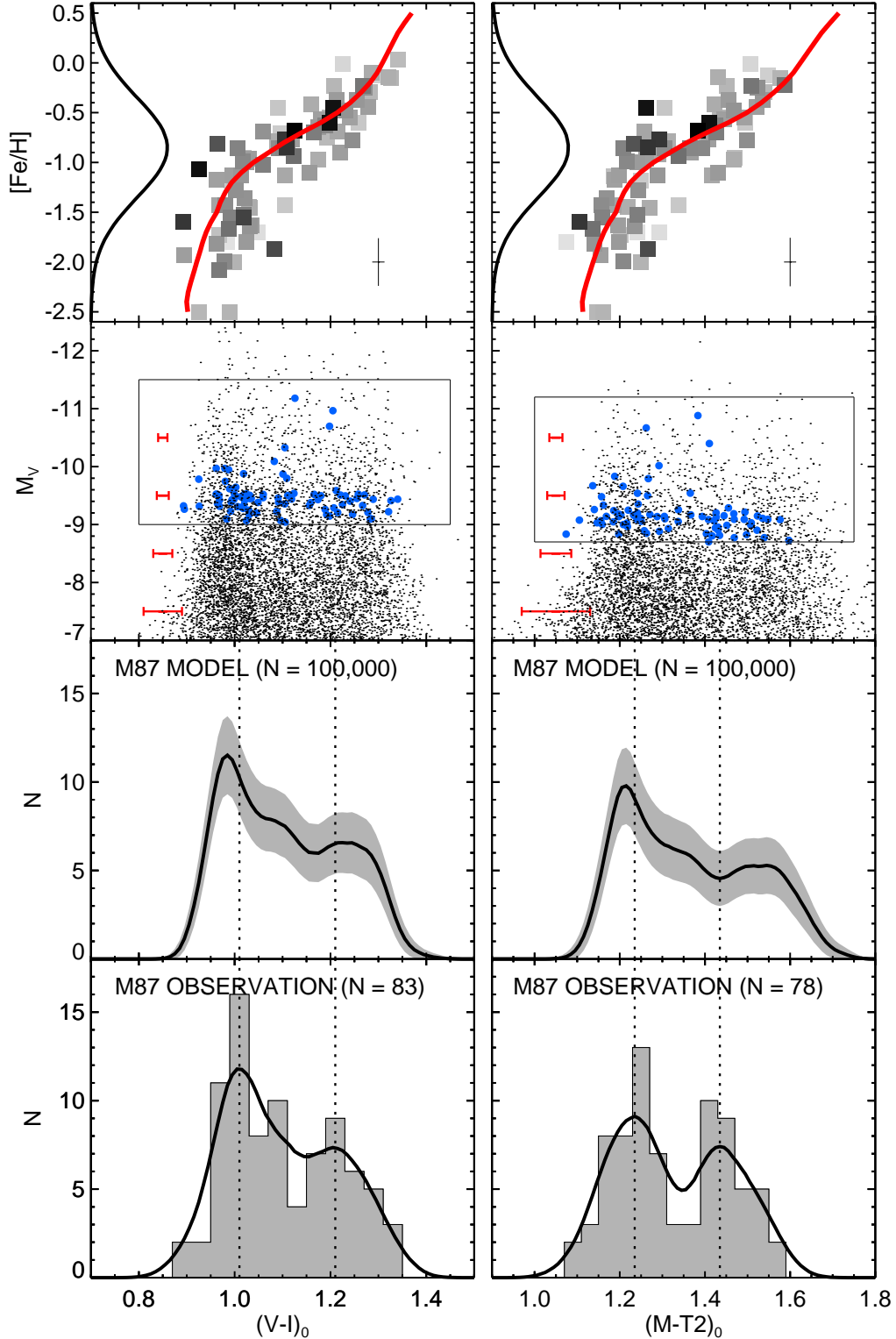}
\caption{\scriptsize
Monte Carlo simulations for the color distributions of M87 GCs. 
(Top row) The color$-$metallicity distributions are compared to the YEPS model predictions of 13.5 Gyr and [$\alpha$/Fe] = 0.3 (red solid lines). 
The error bars show the typical observational errors in [Fe/H] and color.
Along the y-axes, we show single Gaussian MDFs of $10^{6}$ model GCs with the mean [Fe/H] and its standard deviation of $-$0.85 and 0.5.
(Second row) The color$-$magnitude diagrams of $\sim$\,10000 randomly selected model GCs for the $V-I$ and $M-T2$ colors are shown as black dots and the observations as blue circles.
(Third row) Simulated color histograms of GCs, which correspond to the boxed range of the second row, are obtained through the best-fit model lines.
Observational uncertainties are taken into account for the conversion from metallicities to colors.
We randomly draw 83 and 78 GCs from the $10^{6}$ model GCs 10,000 times each for $V-I$ and $M-T2$, and the gray shaded bands show the resultant $1\sigma$ distributions.
The vertical dotted lines are the blue and red peak locations of the observed histogram in the bottom row.
(Bottom row) The observed color distributions are shown.
The solid lines represent the Gaussian kernels for which mean color errors are adopted for the kernel bandwidths.
The vertical dotted lines are the blue and red peak locations.}
\label{M87_15}
\end{center}
\end{figure}

The nonlinear-CMR scenario for GC color bimodality has been tested in the previous series of studies (Papers I\,$\sim$\,IX) using multiple heterogeneous data sets of extragalactic GC systems, both photometric and spectroscopic, and found to be successful in explaining the bimodality phenomenon.
Here we have used homogeneous spectroscopic data of the M87 GC system to continue to examine and test the nonlinear-CMR scenario.
We have shown that M87 GCs exhibit nonlinearity in CMRs (Figure \ref{M87_13}).
We have also shown that the MDF of the M87 GC system is characterized by a broad, unimodal distribution (Figure \ref{M87_14}).
In order to strengthen the statistical significance of the MDF, we have combined the metallicity measurements of our study and those of C98 with high S/N, making one of the largest catalogs of spectroscopically derived metallicity on M87 GCs.
This combined distribution of metallicities derived with the spectroscopic data is consistent with the GC MDFs of several elliptical galaxies inferred from various colors through nonlinear CMRs (Papers II, III, and IV).
Finally, the observed bimodal color distributions are reproduced using our nonlinear metallicity-to-color transformation scheme (Figure \ref{M87_15}), bringing our results in good agreement with the findings of the previous studies.

\begin{figure}[h!]
\begin{center}
\includegraphics[width=16cm]{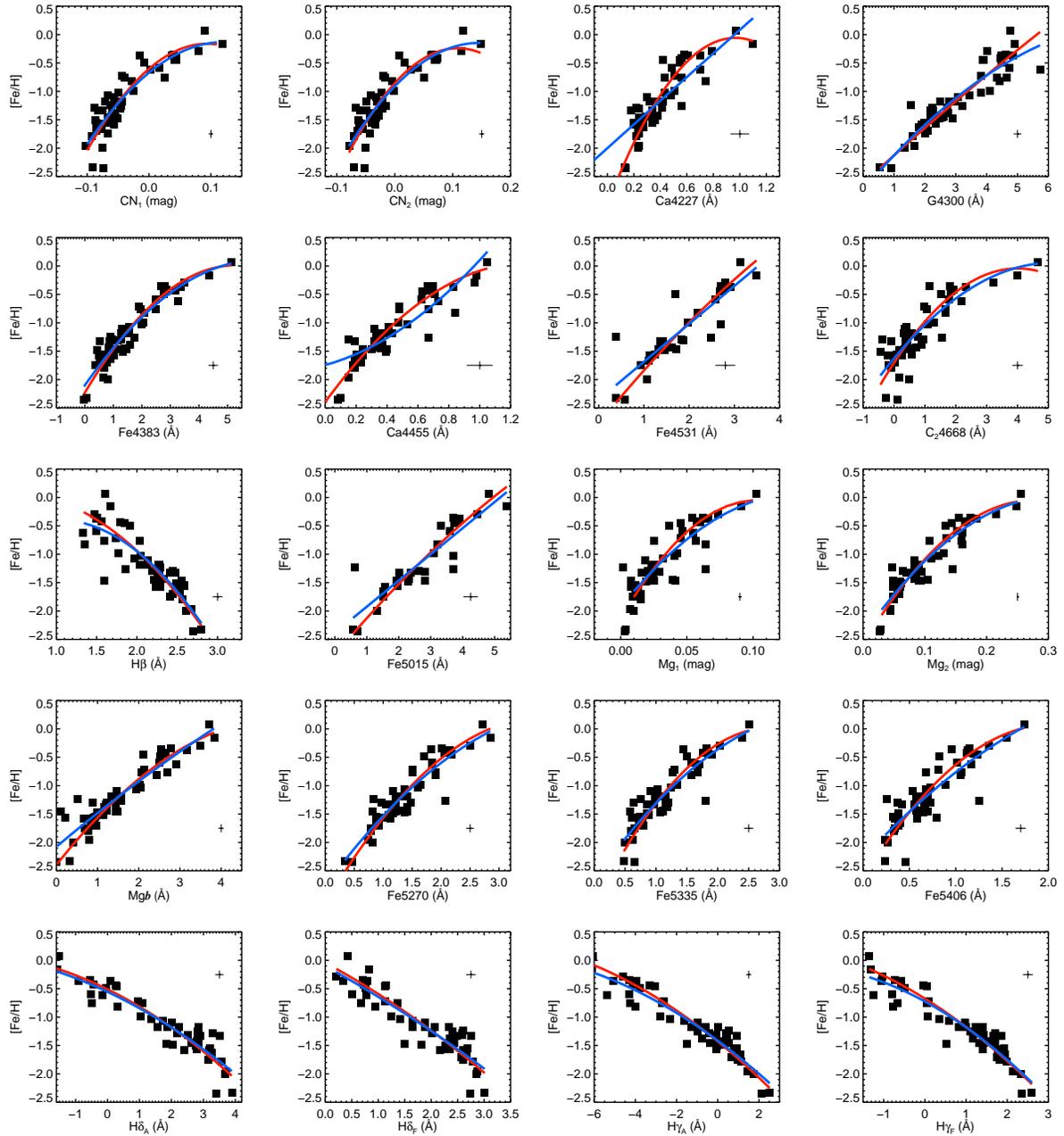}
\caption{Lick indices of 53 MWGCs \citep{kim16} as a function of [Fe/H] \citep{harris1996}. 
The second-order orthogonal distance regression fits to the MWGC data are shown as red lines and the secnd-order least-squares fits are shown as blue lines.
\label{M87_10}}
\end{center}
\end{figure}

Recently, \citet{Villaume19} reported a new metallicity measurement on M87 GCs.
Via full-spectrum SSP model fitting, they obtained a bimodal MDF with systematically higher metallicities compared to the previous findings on M87 (C98, \citet{peng06}).
Their Figure \ref{M87_10} compares the M87 CMR with the literature (\citet{peng06}) and the MWGCs, and they found that the shapes of CMRs differ significantly.
Villaume et al. suggested that the difference in the metal-poor region might be due to the age$-$metallicity degeneracy or the effect of blue horizontal-branch stars.
Given that many extragalactic GCs are found to be old \citep[e.g., C98;][]{Forbes2001, puzia05} and that the YEPS models employ a realistic treatment of horizontal-branch stars in stellar population modeling, we do not consider them major causes for the discrepancy.
We suspect the distinct CMRs might be the result of a different approach to deriving metallicities.
\citet{fahrion2020} compares GC metallicities of the galaxies in the Fornax cluster obtained with different techniques and various sets of constraints in their appendix.
For example, the CMRs obtained by full-spectrum fitting and by using line-strength indices clearly exhibit offsets where GCs have different
metallicities at fixed colors (see their Figure A.1).

In the conventional scenarios of GC formation, blue GCs come from the accretion of low-mass neighboring galaxies.
However, whether the accretion of metal-poor GCs is solely responsible for blue GCs is still an open question.
A piece of evidence against the accretion scenario is the observed sharp blue peaks of giant elliptical galaxies. The mean colors of GC systems correlate tightly with host galaxies’ mass \citep{larsen01, strader2004, peng06}. Such correlation suggests that accreted galaxies used to have a narrow mass range to form the sharp blue peak, which is unlikely.
Further counterevidence is the significant fraction of blue GCs in galaxies in lower-density and isolated regions \citep{peng06, 2012MNRAS.422.3591C}, where galaxies have few accompanying galaxies and thus it is difficult to acquire sufficient metal-poor GCs via accretion. 
Moreover, in our study, a unimodal shape of GC MDF is obtained in the inner main body of M87 ($<$\,6.3 R$_{\rm eff}$), where accreted GCs should be less populated than in the remote halo (see Paper III for further discussion).

Current hierarchical models of galaxy formation predict that the emergence of a massive galaxy involved a large number of protogalaxies, which may leave little room for the existence of merely two GC subpopulations in galaxies such as M87.
In these models, the chemical evolution in the early stage of galaxy formation took place in a rather quasi-monolithic manner, which naturally produces a broad, unimodal MDF of GCs.
We conclude from our results in this study, together with the evidence offered from our series of observational and theoretical studies (Papers I\,$\sim$\,IX), that in the case of massive giant galaxies like M87, GCs are formed on a relatively short timescale via early, vigorous star formation \citep[e.g., Paper III,][]{chiosi14}.

\acknowledgments
{S.-J.Y. acknowledges support from the Mid-career Researcher Program (No. 2019R1A2C3006242) and the SRC Program (the Center for Galaxy Evolution Research; No. 2017R1A5A1070354) through the National Research Foundation of Korea. We would like to thank N. Tamura and T. Hattori for their help and useful discussions.}

\vspace{3cm}
\bibliographystyle{aasjournal}

\begin{thebibliography}{}

\bibitem[{{Ashman} \& {Zepf}(1992)}]{ashman92}
{Ashman}, K.~M., \& {Zepf}, S.~E. 1992, \apj, 384, 50

\bibitem[{{Beasley} {et~al.}(2008){Beasley}, {Bridges}, {Peng}, {Harris},
  {Harris}, {Forbes}, \& {Mackie}}]{beasley08}
{Beasley}, M.~A., {Bridges}, T., {Peng}, E., {et~al.} 2008, \mnras, 386, 1443

\bibitem[{{Blakeslee} {et~al.}(2012){Blakeslee}, {Cho}, {Peng}, {Ferrarese},
  {Jord{\'a}n}, \& {Martel}}]{blakeslee12}
{Blakeslee}, J.~P., {Cho}, H., {Peng}, E.~W., {et~al.} 2012, \apj, 746, 88

\bibitem[{{Brodie} {et~al.}(2012){Brodie}, {Usher}, {Conroy}, {Strader},
  {Arnold}, {Forbes}, \& {Romanowsky}}]{brodie12}
{Brodie}, J.~P., {Usher}, C., {Conroy}, C., {et~al.} 2012, \apjl, 759, L33

\bibitem[{{Bruzual} \& {Charlot}(2003)}]{bruzual03}
{Bruzual}, G., \& {Charlot}, S. 2003, \mnras, 344, 1000

\bibitem[{{Burstein} {et~al.}(1984){Burstein}, {Faber}, {Gaskell}, \&
  {Krumm}}]{burstein84}
{Burstein}, D., {Faber}, S.~M., {Gaskell}, C.~M., \& {Krumm}, N. 1984, \apj,
  287, 586

\bibitem[{{Caldwell} {et~al.}(2011){Caldwell}, {Schiavon}, {Morrison}, {Rose},
  \& {Harding}}]{caldwell11}
{Caldwell}, N., {Schiavon}, R., {Morrison}, H., {Rose}, J.~A., \& {Harding}, P.
  2011, \aj, 141, 61

\bibitem[{{Cardiel} {et~al.}(1998){Cardiel}, {Gorgas}, {Cenarro}, \&
  {Gonzalez}}]{cardiel98}
{Cardiel}, N., {Gorgas}, J., {Cenarro}, J., \& {Gonzalez}, J.~J. 1998, \aaps,
  127, 597

\bibitem[{{Carretta} {et~al.}(2008){Carretta}, {Bragaglia}, {Gratton}, \&
{Lucatello}}]{carretta08}
{Carretta}, E., {Bragaglia}, A., {Gratton}, R.~G., \& {Lucatello}, S. 2008, arXiv:0811.3591

\bibitem[{{Caso} {et~al.}(2017){Caso}, {Bassino}, \& {G{\'o}mez}}]{caso2017}
{Caso}, J.~P., {Bassino}, L.~P., \& {G{\'o}mez}, M. 2017, \mnras, 470, 3227

\bibitem[{{Cenarro} {et~al.}(2007){Cenarro}, {Beasley}, {Strader}, {Brodie}, \&
  {Forbes}}]{cenarro07}
{Cenarro}, A.~J., {Beasley}, M.~A., {Strader}, J., {Brodie}, J.~P., \&
  {Forbes}, D.~A. 2007, \aj, 134, 391

\bibitem[{{Chies-Santos} {et~al.}(2012){Chies-Santos}, {Larsen}, {Cantiello},
  {Strader}, {Kuntschner}, {Wehner}, \& {Brodie}}]{chies12}
{Chies-Santos}, A.~L., {Larsen}, S.~S., {Cantiello}, M., {et~al.} 2012, \aap,
  539, A54

\bibitem[{{Chiosi} {et~al.}(2014){Chiosi}, {Merlin}, {Piovan}, \&
  {Tantalo}}]{chiosi14}
{Chiosi}, C., {Merlin}, E., {Piovan}, L., \& {Tantalo}, R. 2014, Galaxies, 2,
  300

\bibitem[{{Cho} {et~al.}(2016){Cho}, {Blakeslee}, {Chies-Santos}, {Jee},
  {Jensen}, {Peng}, \& {Lee}}]{cho16}
{Cho}, H., {Blakeslee}, J.~P., {Chies-Santos}, A.~L., {et~al.} 2016, \apj, 822,
  95

\bibitem[Cho et al.(2012)]{2012MNRAS.422.3591C} Cho, J., Sharples, R.~M., Blakeslee, J.~P., et al.\ 2012, \mnras, 422, 3591

\bibitem[{{Choksi} {et~al.}(2018){Choksi}, {Gnedin}, \& {Li}}]{choksi2018}
{Choksi}, N., {Gnedin}, O.~Y., \& {Li}, H. 2018, \mnras, 480, 2343

\bibitem[{{Chung} {et~al.}(2013){Chung}, {Yoon}, {Lee}, \& {Lee}}]{chung13}
{Chung}, C., {Yoon}, S.-J., {Lee}, S.-Y., \& {Lee}, Y.-W. 2013, \apjs, 204, 3

\bibitem[Chung et al.(2016)]{chung2016} Chung, C., Yoon, S.-J., Lee, S.-Y., et al.\ 2016, \apj, 818, 201

\bibitem[{Chung {et~al.}(2017)Chung, Yoon, \& Lee}]{chung17}
Chung, C., Yoon, S.-J., \& Lee, Y.-W. 2017, \apj, 842, 91

\bibitem[{{Cohen} {et~al.}(1998){Cohen}, {Blakeslee}, \& {Ryzhov}}]{cohen98}
{Cohen}, J.~G., {Blakeslee}, J.~P., \& {Ryzhov}, A. 1998, \apj, 496, 808

\bibitem[{{Cote} {et~al.}(1998){Cote}, {Marzke}, \& {West}}]{cote98}
{Cote}, P., {Marzke}, R.~O., \& {West}, M.~J. 1998, \apj, 501, 554

\bibitem[{{Cottrell} \& {Da Costa}(1981)}]{cottrell81}
{Cottrell}, P.~L., \& {Da Costa}, G.~S. 1981, \apjl, 245, L79

\bibitem[{{D'Antona} \& {Ventura}(2007)}]{dantona07}
{D'Antona}, F., \& {Ventura}, P. 2007, \mnras, 379, 1431

\bibitem[{{De B{\'o}rtoli} {et~al.}(2020){De B{\'o}rtoli}, {Bassino}, {Caso}, \& {Ennis}}]{debortoli20} {De B{\'o}rtoli}, B.~J., {Bassino}, L.~P., {Caso}, J.~P., \& {Ennis}, A.~I. 2020, \mnras, 492, 4313

\bibitem[{{Decressin}{et~al.}(2007){Meynet}, {Charbonnel}, {Prantzos}, \& {Ekstr{\"o}m}}]{decressin2007}
{Decressin}, T, {Meynet}, G, {Charbonnel}, C, {Prantzos}, N, {Ekstr{\"o}m}, S. 2007, \aap, 464, 3

\bibitem[{{D'Ercole} {et~al.}(2008){D'Ercole}, {Vesperini}, {D'Antona},
  {McMillan}, \& {Recchi}}]{dercole08}
{D'Ercole}, A., {Vesperini}, E., {D'Antona}, F., {McMillan}, S.~L.~W., \&
  {Recchi}, S. 2008, \mnras, 391, 825

\bibitem[{{Dirsch} {et~al.}(2003){Dirsch}, {Richtler}, {Geisler}, {Forte},
  {Bassino}, \& {Gieren}}]{dirsch03}
{Dirsch}, B., {Richtler}, T., {Geisler}, D., {et~al.} 2003, \aj, 125, 1908

\bibitem[{{Ennis} {et~al.}(2019){Ennis}, {Bassino}, {Caso}, \& {De B{\'o}rtoli}}]{ennis2019}
{Ennis}, A.~I., {Bassino}, L.~P., {Caso}, J.~P., \& {De B{\'o}rtoli}, B.~J. 2019, \mnras, 488, 770

\bibitem[{{Faber} {et~al.}(1985){Faber}, {Friel}, {Burstein}, \& {Gaskell}}]{faber85}
{Faber}, S.~M., {Friel}, E.~D., {Burstein}, D., \& {Gaskell}, C.~M. 1985, \apjs, 57, 711

\bibitem[{{Fahrion} {et~al.}(2020){Fahrion}, {Lyubenova}, {Hilker}, {van de
  Ven}, {Falc{\'o}n-Barroso}, {Leaman}, {Mart{\'\i}n-Navarro}, {Bittner},
  {Coccato}, {Corsini}, {Gadotti}, {Iodice}, {McDermid}, {Pinna}, {Sarzi},
  {Viaene}, {de Zeeuw}, \& {Zhu}}]{fahrion2020}
{Fahrion}, K., {Lyubenova}, M., {Hilker}, M., {et~al.} 2020, \aap, 637, A27

\bibitem[{{Forbes} {et~al.}(2001){Forbes}, {Beasley}, {Brodie}, \&
  {Kissler-Patig}}]{Forbes2001}
{Forbes}, D.~A., {Beasley}, M.~A., {Brodie}, J.~P., \& {Kissler-Patig}, M.
  2001, \apjl, 563, L143

\bibitem[{{Forbes} {et~al.}(1997){Forbes}, {Brodie}, \& {Grillmair}}]{forbes97}
{Forbes}, D.~A., {Brodie}, J.~P., \& {Grillmair}, C.~J. 1997, \aj, 113, 1652

\bibitem[{{Forbes} \& {Remus}(2018)}]{forbes2018}
{Forbes}, D.~A., \& {Remus}, R.-S. 2018, \mnras, 479, 4760

\bibitem[{{Foster} {et~al.}(2010){Foster}, {Forbes}, {Proctor}, {Strader},
  {Brodie}, \& {Spitler}}]{foster10}
{Foster}, C., {Forbes}, D.~A., {Proctor}, R.~N., {et~al.} 2010, \aj, 139, 1566


\bibitem[{{Geisler} {et~al.}(1996){Geisler}, {Lee}, \& {Kim}}]{geisler96}
{Geisler}, D., {Lee}, M.~G., \& {Kim}, E. 1996, \aj, 111, 1529

\bibitem[{{Gerber} {et~al.}(2018){Gerber}, {Friel}, \& {Vesperini}}]{gerber2018}
{Gerber}, Jeffrey M., {Friel}, Eileen D., {Vesperini}, Enrico. 2018, \aj, 156, 6G


\bibitem[{{Gorgas} {et~al.}(1993){Gorgas}, {Faber}, {Burstein}, {Gonzalez}, {Courteau}, \& {Prosser}}]{gorgas1993}
{Gorgas}, J., {Faber}, S.~M., {Burstein}, D., {Gonzalez}, J.~J., {Courteau}, S., \& {Prosser}, C. 1993, \apjs, 86, 153


\bibitem[{{Hanes} {et~al.}(2001){Hanes}, {C{\^o}t{\'e}}, {Bridges},
  {McLaughlin}, {Geisler}, {Harris}, {Hesser}, \& {Lee}}]{hanes01}
{Hanes}, D.~A., {C{\^o}t{\'e}}, P., {Bridges}, T.~J., {et~al.} 2001, \apj, 559, 812

\bibitem[{{Harris}(1996)}]{harris1996}
{Harris}, W.~E. 1996, \aj, 112, 1487

\bibitem[{{Harris} {et~al.}(2017){Harris}, {Ciccone}, {Eadie}, {Gnedin},
  {Geisler}, {Rothberg}, \& {Bailin}}]{harris2017}
{Harris}, W.~E., {Ciccone}, S.~M., {Eadie}, G.~M., {et~al.} 2017, \apj, 835,
  101

\bibitem[{{Harris} {et~al.}(2009){Harris}, {Kavelaars}, {Hanes}, {Pritchet}, \&
  {Baum}}]{harris09}
{Harris}, W.~E., {Kavelaars}, J.~J., {Hanes}, D.~A., {Pritchet}, C.~J., \&
  {Baum}, W.~A. 2009, \aj, 137, 3314

\bibitem[{{Harris} {et~al.}(2006){Harris}, {Whitmore}, {Karakla}, {Oko{\'n}},
  {Baum}, {Hanes}, \& {Kavelaars}}]{harris06}
{Harris}, W.~E., {Whitmore}, B.~C., {Karakla}, D., {et~al.} 2006, \apj, 636, 90

\bibitem[{{Jefferys} {et~al.}(1988){Jefferys}, {Fitzpatrick}, \&
  {McArthur}}]{jefferys88}
{Jefferys}, W.~H., {Fitzpatrick}, M.~J., \& {McArthur}, B.~E. 1988, Celestial
  Mechanics, 41, 39

\bibitem[{{Jord{\'a}n} {et~al.}(2009){Jord{\'a}n}, {Peng}, {Blakeslee},
  {C{\^o}t{\'e}}, {Eyheramendy}, {Ferrarese}, {Mei}, {Tonry}, \&
  {West}}]{jordan09}
{Jord{\'a}n}, A., {Peng}, E.~W., {Blakeslee}, J.~P., {et~al.} 2009, \apjs, 180,
  54

\bibitem[{{Kartha} {et~al.}(2014){Kartha}, {Forbes}, {Spitler}, {Romanowsky},
  {Arnold}, \& {Brodie}}]{kartha14}
{Kartha}, S.~S., {Forbes}, D.~A., {Spitler}, L.~R., {et~al.} 2014, \mnras, 437,
  273

\bibitem[{{Kartha} {et~al.}(2016){Kartha}, {Forbes}, {Alabi}, {Brodie},
  {Romanowsky}, {Strader}, {Spitler}, {Jennings}, \& {Roediger}}]{kartha2016}
{Kartha}, S.~S., {Forbes}, D.~A., {Alabi}, A.~B., {et~al.} 2016, \mnras, 458,
  105


\bibitem[{{Kim} {et~al.}(2013){Kim}, {Yoon}, {Chung}, {Caldwell}, {Schiavon},
  {Kang}, {Rey}, \& {Lee}}]{kim13}
{Kim}, S., {Yoon}, S.-J., {Chung}, C., {et~al.} 2013, \apj, 768, 138


\bibitem[{Kim {et~al.}(2016)Kim, Cho, Sharples, Vazdekis, Beasley, \&
  Yoon}]{kim16}
Kim, H.-S., Cho, J., Sharples, R.~M., {et~al.} 2016, The Astrophysical Journal
  Supplement Series, 227, 24

\bibitem[{{Kim} \& {Yoon}(2017)}]{kim17}
{Kim}, S., \& {Yoon}, S.-J. 2017, \apj, 843, 43


\bibitem[{{Kim} \& {Lee}(2018)}]{kim2018}
{Kim}, Jenny J., {Lee}, Young-Wook. 2018, \apj, 869, 35

\bibitem[{{Kundu} \& {Whitmore}(2001)}]{kundu01}
{Kundu}, A., \& {Whitmore}, B.~C. 2001, \aj, 121, 2950


\bibitem[{{Kuntschner} {et~al.}(2010){Kuntschner}, {Emsellem}, {Bacon},
  {Cappellari}, {Davies}, {de Zeeuw}, {Falc{\'o}n-Barroso}, {Krajnovi{\'c}},
  {McDermid}, {Peletier}, {Sarzi}, {Shapiro}, {van den Bosch}, \& {van de
  Ven}}]{kuntschner2010}
{Kuntschner}, H., {Emsellem}, E., {Bacon}, R., {et~al.} 2010, \mnras, 408, 97

\bibitem[{{Landolt}(1992)}]{landolt1992}
{Landolt}, Arlo U. 1992, \aj, 102, 340

\bibitem[{{Larsen} {et~al.}(2001){Larsen}, {Brodie}, {Huchra}, {Forbes}, \&
  {Grillmair}}]{larsen01}
{Larsen}, S.~S., {Brodie}, J.~P., {Huchra}, J.~P., {Forbes}, D.~A., \&
  {Grillmair}, C.~J. 2001, \aj, 121, 2974

\bibitem[{{Lee} {et~al.}(2010){Lee}, {Park}, {Hwang}, {Arimoto}, {Tamura}, \&
  {Onodera}}]{lee10}
{Lee}, M.~G., {Park}, H.~S., {Hwang}, H.~S., {et~al.} 2010, \apj, 709, 1083

\bibitem[{{Lee} {et~al.}(2008){Lee}, {Hwang}, {Park}, {Park}, {Kim}, {Sohn},
  {Lee}, {Rey}, {Lee}, \& {Kim}}]{lee08}
{Lee}, M.~G., {Hwang}, H.~S., {Park}, H.~S., {et~al.} 2008, \apj, 674, 857

\bibitem[{{Lee}(2016)}]{lee2016}
{Lee}, Jae-Woo. 2016, \apjs, 226, 16L


\bibitem[{{Lee} {et~al.}(2019){Lee}, {Chung}, \& {Yoon}}]{sylee2019}
{Lee}, S.-Y., {Chung}, C., \& {Yoon}, S.-J. 2019, \apjs, 240, 2

\bibitem[Lee et al.(2020)]{sylee2020} Lee, S.-Y., Chung, C., \& Yoon, S.-J.\ 2020, \apj, 905, 124


\bibitem[{{Liu} {et~al.}(2011){Liu}, {Peng}, {Jord{\'a}n}, {Ferrarese},
  {Blakeslee}, {C{\^o}t{\'e}}, \& {Mei}}]{liu11}
{Liu}, C., {Peng}, E.~W., {Jord{\'a}n}, A., {et~al.} 2011, \apj, 728, 116


\bibitem[{{MacLean} {et~al.}(2018)}]{macLean2018}
{MacLean}, B.~T., {Campbell}, S.~W., {Amarsi}, A.~M., et al. 2018, \mnras, 481, 373


\bibitem[{{Maeder} \& {Meynet}(2006)}]{maeder06}
{Maeder}, A., \& {Meynet}, G. 2006, \aap, 448, L37

\bibitem[{{Matteucci}(1994)}]{matteucci94}
{Matteucci}, F. 1994, \aap, 288, 57

\bibitem[{{Mieske} {et~al.}(2006){Mieske}, {Jord{\'a}n}, {C{\^o}t{\'e}},
  {Kissler-Patig}, {Peng}, {Ferrarese}, {Blakeslee}, {Mei}, {Merritt}, {Tonry},
  \& {West}}]{mieske06}
{Mieske}, S., {Jord{\'a}n}, A., {C{\^o}t{\'e}}, P., {et~al.} 2006, \apj, 653,
  193

\bibitem[Mieske et al.(2010)]{mieske10} Mieske, S., Jord{\'a}n, A., C{\^o}t{\'e}, P., et al.\ 2010, \apj, 710, 1672

\bibitem[{{Monet} {et~al.}(2003)}]{monet2003}
{Monet}, David G., {Levine}, Stephen E., {Canzian}, Blaise., et al. 2003, \aj, 125, 984


\bibitem[{{Mucciarelli} {et~al.}(2015){Mucciarelli}, {Lapenna}, {Massari}, {Pancino},  {Stetson}, {Ferraro}, {Lanzoni}, \& {Lardo}}]{mucciarelli2015}
{Mucciarelli}, A., {Lapenna}, E., {Massari}, D., {Pancino}, E., {Stetson}, P.~B., {Ferraro}, F.~R., {Lanzoni}, B., \& {Lardo}, C. 2015, \apj, 809, 128


\bibitem[{{Ostrov} {et~al.}(1993){Ostrov}, {Geisler}, \& {Forte}}]{ostrov93}
{Ostrov}, P., {Geisler}, D., \& {Forte}, J.~C. 1993, \aj, 105, 1762

\bibitem[{{Peng} {et~al.}(2004){Peng}, {Ford}, \& {Freeman}}]{peng04}
{Peng}, E.~W., {Ford}, H.~C., \& {Freeman}, K.~C. 2004, \apj, 602, 705

\bibitem[{{Peng} {et~al.}(2006){Peng}, {Jord{\'a}n}, {C{\^o}t{\'e}},
  {Blakeslee}, {Ferrarese}, {Mei}, {West}, {Merritt}, {Milosavljevi{\'c}}, \&
  {Tonry}}]{peng06}
{Peng}, E.~W., {Jord{\'a}n}, A., {C{\^o}t{\'e}}, P., {et~al.} 2006, \apj, 639,
  95

\bibitem[{{Pota} {et~al.}(2013){Pota}, {Forbes}, {Romanowsky}, {Brodie},
  {Spitler}, {Strader}, {Foster}, {Arnold}, {Benson}, {Blom}, {Hargis},
  {Rhode}, \& {Usher}}]{pota13}
{Pota}, V., {Forbes}, D.~A., {Romanowsky}, A.~J., {et~al.} 2013, \mnras, 428,
  389

\bibitem[{{Prantzos} \& {Charbonnel}(2006)}]{prantzos06}
{Prantzos}, N., \& {Charbonnel}, C. 2006, \aap, 458, 135

\bibitem[{{Proctor} {et~al.}(2004){Proctor}, {Forbes}, \&
  {Beasley}}]{proctor04}
{Proctor}, R.~N., {Forbes}, D.~A., \& {Beasley}, M.~A. 2004, \mnras, 355, 1327

\bibitem[{{Puzia} {et~al.}(2005){Puzia}, {Perrett}, \& {Bridges}}]{puzia05}
{Puzia}, T.~H., {Perrett}, K.~M., \& {Bridges}, T.~J. 2005, \aap, 434, 909

\bibitem[{{Renzini}(2008)}]{renzini2008}
{Renzini}, Alvio. 2008, \mnras, 391, 354

\bibitem[{{Richtler}(2006)}]{richtler06}
{Richtler}, T. 2006, Bulletin of the Astronomical Society of India, 34, 83

\bibitem[{{Richtler} {et~al.}(2015){Richtler}, {Salinas}, {Lane}, {Hilker}, \&
  {Schirmer}}]{richtler15}
{Richtler}, T., {Salinas}, R., {Lane}, R.~R., {Hilker}, M., \& {Schirmer}, M.
  2015, \aap, 574, A21

\bibitem[{{Salpeter}(1955)}]{Salpeter1955}
{Salpeter}, E. E. 1955, \apj, 121, 161S

\bibitem[{{Schlegel} {et~al.}(1998){Finkbeiner}, \& {Davis}}]{schlegel1998}
{Schlegel}, David J., {Finkbeiner}, Douglas P., \& {Davis}, Marc. 1998, \apj, 500, 525

\bibitem[{{Sinnott} {et~al.}(2010){Sinnott}, {Hou}, {Anderson}, {Harris}, \&
  {Woodley}}]{sinnott10}
{Sinnott}, B., {Hou}, A., {Anderson}, R., {Harris}, W.~E., \& {Woodley}, K.~A.
  2010, \aj, 140, 2101
  
 \bibitem[{{Stetson}(1987)}]{stetson1987}
{Stetson}, Peter B. 1987, \pasp, 99, 191

\bibitem[{{Stetson}(2000)}]{stetson2000}
{Stetson}, Peter B. 2000, \pasp, 112, 925

\bibitem[{{Strader} \& {Brodie}(2004)}]{strader2004}
{Strader}, J., \& {Brodie}, J.~P. 2004, \aj, 128, 1671

\bibitem[{{Strader} {et~al.}(2006){Strader}, {Brodie}, {Spitler}, \&
  {Beasley}}]{strader06}
{Strader}, J., {Brodie}, J.~P., {Spitler}, L., \& {Beasley}, M.~A. 2006, \aj,
  132, 2333

\bibitem[{{Strader} {et~al.}(2011){Strader}, {Romanowsky}, {Brodie}, {Spitler},
  {Beasley}, {Arnold}, {Tamura}, {Sharples}, \& {Arimoto}}]{strader11}
{Strader}, J., {Romanowsky}, A.~J., {Brodie}, J.~P., {et~al.} 2011, \apjs, 197,
  33

\bibitem[{{Tamura} {et~al.}(2006{\natexlab{a}}){Tamura}, {Sharples}, {Arimoto},
  {Onodera}, {Ohta}, \& {Yamada}}]{tamura06a}
{Tamura}, N., {Sharples}, R.~M., {Arimoto}, N., {et~al.} 2006{\natexlab{a}},
  \mnras, 373, 588

\bibitem[Tamura et al.(2006{\natexlab{b}})]{tamura06b} Tamura, N., Sharples, R.~M., Arimoto, N., et al.\ 2006{\natexlab{b}}, \mnras, 373, 601

\bibitem[{{Thomas} {et~al.}(2003){Thomas}, {Maraston}, \& {Bender}}]{thomas03}
{Thomas}, D., {Maraston}, C., \& {Bender}, R. 2003, \mnras, 339, 897

\bibitem[{{Thomas} {et~al.}(2005){Thomas}, {Maraston}, {Bender}, \& {Mendes de
  Oliveira}}]{thomas05}
{Thomas}, D., {Maraston}, C., {Bender}, R., \& {Mendes de Oliveira}, C. 2005,
  \apj, 621, 673

\bibitem[{{Tonry} \& {Davis}(1979)}]{tonry79}
{Tonry}, J., \& {Davis}, M. 1979, \aj, 84, 1511

\bibitem[{{Trager} {et~al.}(2000){Trager}, {Faber}, {Worthey}, \&
  {Gonz{\'a}lez}}]{trager00}
{Trager}, S.~C., {Faber}, S.~M., {Worthey}, G., \& {Gonz{\'a}lez}, J.~J. 2000,
  \aj, 120, 165

\bibitem[{{Trager} {et~al.}(1998){Trager}, {Worthey}, {Faber}, {Burstein}, \&
  {Gonzalez}}]{trager98}
{Trager}, S.~C., {Worthey}, G., {Faber}, S.~M., {Burstein}, D., \& {Gonzalez},
  J.~J. 1998, \apjs, 116, 1

\bibitem[{{Tripicco} \& {Bell}(1995)}]{tripicco95}
{Tripicco}, M.~J., \& {Bell}, R.~A. 1995, \aj, 110, 3035

\bibitem[{{Usher} {et~al.}(2015){Usher}, {Forbes}, {Brodie}, {Romanowsky},
  {Strader}, {Conroy}, {Foster}, {Pastorello}, {Pota}, \& {Arnold}}]{usher15}
{Usher}, C., {Forbes}, D.~A., {Brodie}, J.~P., {et~al.} 2015, \mnras, 446, 369

\bibitem[{{Valdes}(1998)}]{valdes1998}{Valdes}, F.~G. 1998, in ASP Conf. Ser. 145, Astronomical Data Analysis Software and Systems VII, ed. R. Albrecht, R. N. Hook, \& H. A. Bushouse (San Francisco, CA: ASP), 53

\bibitem[{{Villaume} {et~al.}(2019){Villaume}, {Romanowsky}, {Brodie}, \&
  {Strader}}]{Villaume19}
{Villaume}, A., {Romanowsky}, A.~J., {Brodie}, J., \& {Strader}, J. 2019, \apj,
  879, 45

\bibitem[{{Woodley} {et~al.}(2010){Woodley}, {Harris}, {Puzia}, {G{\'o}mez},
  {Harris}, \& {Geisler}}]{woodley10}
{Woodley}, K.~A., {Harris}, W.~E., {Puzia}, T.~H., {et~al.} 2010, \apj, 708,
  1335

\bibitem[{{Worthey}(1994)}]{worthey94}
{Worthey}, G. 1994, \apjs, 95, 107

\bibitem[{{Worthey} {et~al.}(1992){Worthey}, {Faber}, \&
  {Gonzalez}}]{worthey92}
{Worthey}, G., {Faber}, S.~M., \& {Gonzalez}, J.~J. 1992, \apj, 398, 69

\bibitem[{{Worthey} \& {Ottaviani}(1997)}]{worthey97}
{Worthey}, G., \& {Ottaviani}, D.~L. 1997, \apjs, 111, 377

\bibitem[{{Yoon} {et~al.}(2013){Yoon}, {Sohn}, {Kim}, {Chung}, {Cho}, {Lee}, \&
  {Blakeslee}}]{yoon13}
{Yoon}, S.-J., {Sohn}, S.~T., {Kim}, H.-S., {et~al.} 2013, \apj, 768, 137

\bibitem[{{Yoon} {et~al.}(2011{\natexlab{a}}){Yoon}, {Sohn}, {Lee}, {Kim},
  {Cho}, {Chung}, \& {Blakeslee}}]{yoon11a}
{Yoon}, S.-J., {Sohn}, S.~T., {Lee}, S.-Y., {et~al.} 2011{\natexlab{a}}, \apj,
  743, 149

\bibitem[{{Yoon} {et~al.}(2006){Yoon}, {Yi}, \& {Lee}}]{yoon06}
{Yoon}, S.-J., {Yi}, S.~K., \& {Lee}, Y.-W. 2006, Science, 311, 1129

\bibitem[{{Yoon} {et~al.}(2011{\natexlab{b}}){Yoon}, {Lee}, {Blakeslee},
  {Peng}, {Sohn}, {Cho}, {Kim}, {Chung}, {Kim}, \& {Lee}}]{yoon11b}
{Yoon}, S.-J., {Lee}, S.-Y., {Blakeslee}, J.~P., {et~al.} 2011{\natexlab{b}},
  \apj, 743, 150

\bibitem[{{Zinn} \& {West}(1984)}]{zinn1984}
{Zinn}, R., \& {West}, M.~J. 1984, \apjs, 55, 45

\end{thebibliography}

\clearpage

\clearpage

\clearpage
\begin{table}
\begin{center}
\caption{FOCAS Observing Log\label{tbl1}}
\begin{tabular}{ccc}
\hline
\hline
\multicolumn{1}{c}{Field} &
\multicolumn{1}{c}{Exposure Time} &
\multicolumn{1}{c}{N of slits}  \\
\hline
1 & 9.9 hr &  43 \\
2 & 5.5 hr &  40 \\
3 & 6.0 hr &  46 \\
5 & 1.0 hr &  28 \\
\hline
\end{tabular}
\vspace{0.3cm}
\end{center}
\end{table}

\begin{table}
\begin{center}
\caption{Radial velocity measurements of M87 GCs.\label{tbl2}}
\begin{tabular}{cccccc}
\hline
\hline
\multicolumn{1}{c}{ID} & 
\multicolumn{1}{c}{R.A. (J2000)} & 
\multicolumn{1}{c}{Dec. (J2000)} & 
\multicolumn{1}{c}{$V$ (mag)} & 
\multicolumn{1}{c}{$V-I$ (mag)} & 
\multicolumn{1}{c}{Velocity (km\,s$^{-1}$)} \\
\hline
GC100	&	       12.51477 &	12.45610 	&	21.85 &	1.24 &	1293.72 \\ 
GC101	&	       12.51464 &	12.45117 	&	22.27 &	1.05 &	1188.20 \\ 
GC102	&	       12.51483 &	12.45163 	&	21.80 &	1.08 &	1449.52 \\ 
GC103	&	       12.51649 &	12.47256 	&	21.88 &	1.19 &	1208.59 \\ 
GC104	&	       12.51627 &	12.46610 	&	21.77 &	1.18 &	1648.94 \\ 
GC105	&	       12.51435 &	12.43381 	&	21.50 &	0.96 &	1939.86 \\ 
GC106	&	       12.51635 &	12.46093 	&	21.49 &	1.12 &	1008.90 \\ 
GC107	&	       12.51515 &	12.44067 	&	21.50 &	1.16 &	1422.68 \\ 
GC108	&	       12.51470 &	12.42985 	&	21.35 &	1.32 &	1327.10 \\ 
GC111	&	       12.51642 &	12.44943 	&	21.51 &	1.01 &	1188.34 \\ 
\hline
\end{tabular}
\vspace{0.3cm}
\tablecomments{The full table is available in the online version.}
\end{center}
\end{table}

\clearpage
\begin{table}
\tiny
\begin{center}
\caption{Lick indices for GCs in M87.\label{tbl3}}
\begin{tabular}{ccccccccccccccc}
\hline
\hline
\multicolumn{1}{c}{ID} & 
\multicolumn{1}{c}{Fe4383} &\multicolumn{1}{c}{Ca4455}&\multicolumn{1}{c}{Fe4531} &\multicolumn{1}{c}{C$_2$4668}&\multicolumn{1}{c}{H$\beta$} &\multicolumn{1}{c}{Fe5015}&\multicolumn{1}{c}{Mg$_1$} &\multicolumn{1}{c}{Mg$_2$} &\multicolumn{1}{c}{Mg{\it b}} &\multicolumn{1}{c}{Fe5270}&\multicolumn{1}{c}{Fe5335} &\multicolumn{1}{c}{Fe5406}&\multicolumn{1}{c}{[Fe/H]}&\multicolumn{1}{c}{[M/H]} \\

\multicolumn{1}{c}{ }         &\multicolumn{1}{c}{({\AA})}  &\multicolumn{1}{c}{({\AA})} &\multicolumn{1}{c}{({\AA})}  &\multicolumn{1}{c}{({\AA})} &\multicolumn{1}{c}{({\AA})}  &\multicolumn{1}{c}{({\AA})} &\multicolumn{1}{c}{(mag)}&\multicolumn{1}{c}{(mag)}&\multicolumn{1}{c}{({\AA})}&\multicolumn{1}{c}{({\AA})} &\multicolumn{1}{c}{({\AA})}  &\multicolumn{1}{c}{({\AA})} &\multicolumn{1}{c}{(dex)} &\multicolumn{1}{c}{(dex)}     \\

\hline
GC100	&  1.8615  &  1.0645  &  3.3533  &  3.5297  &  1.7107  &  3.7819  &   0.0727  &  0.2234  &  2.9517  &  2.2039  &   0.6177  &  1.1855 &  -0.7800 &  -0.4694  \\
GC101	& -0.6550  &  1.1529  &  4.0633  &  2.6521  &  1.4611  &  2.8635  &   0.0514  &  0.1146  &  0.3783  &  1.7925  &   0.0719  &  0.7161 &   ...      &   ...   \\
GC102	&  3.8148  &  1.6855  &  3.3981  &  1.3580  &  2.4850  &  3.6432  &   0.0205  &  0.1252  &  2.4836  &  0.8816  &   1.9573  &  0.8608 &  -0.7900 &  -0.7292  \\
GC103	&  3.2314  &  1.8435  &  4.0112  &  6.1488  &  1.4165  &  2.4123  &   0.0591  &  0.2258  &  2.2700  &  1.4950  &   2.4464  &  1.2420 &  -0.5500 &  -0.4117  \\
GC104	&  1.8406  &  1.1563  &  3.0876  &  2.5781  &  1.8860  &  3.4358  &   0.0692  &  0.2112  &  3.1684  &  1.2993  &   1.9182  &  1.5338 &  -0.6300 &  -0.5202  \\
GC105	&  0.6021  & -0.1490  &  1.2060  &  3.2521  &  2.9977  &  0.6581  &  -0.0084  &  0.0424  &  0.8429  &  0.7764  &   0.5759  &  0.3830 &  -1.8200 &  -1.9267  \\
GC106	&  1.7641  &  0.4162  &  2.8021  &  0.8869  &  2.5605  &  3.5117  &   0.0440  &  0.1453  &  2.6785  &  1.6195  &   0.6513  &  1.2566 &  -0.9400 &  -0.8706  \\
GC107	&  2.4128  &  0.9624  &  3.0722  &  4.6723  &  2.1401  &  3.3260  &   0.0580  &  0.1854  &  2.8042  &  2.4059  &   1.7382  &  0.8414 &  -0.7000 &  -0.5606  \\
GC108	&  3.0802  &  1.5461  &  3.1543  &  4.9425  &  2.1846  &  4.9087  &   0.1286  &  0.3280  &  5.6924  &  2.9285  &   2.4460  &  1.4990 &  -0.2200 &  -0.0706  \\
GC111	&  1.7686  &  0.5836  &  1.7412  &  0.9927  &  1.9150  &  3.5745  &   0.0081  &  0.0865  &  1.2262  &  1.1767  &   0.5783  &  0.4703 &  -1.5500 &  -1.2387  \\
\hline
\end{tabular}
\tablecomments{The full table is available in the online version.}
\vspace{0.3cm}
\end{center}
\end{table}

\clearpage
\begin{table}
\tiny
\begin{center}
\caption{Lick index measurement uncertainties for M87 GCs \label{tbl4}}
\begin{tabular}{ccccccccccccc}
\hline
\hline
\multicolumn{1}{c}{ID} & 
\multicolumn{1}{c}{Fe4383} &\multicolumn{1}{c}{Ca4455}&\multicolumn{1}{c}{Fe4531} &\multicolumn{1}{c}{C$_2$4668}&\multicolumn{1}{c}{H$\beta$} &\multicolumn{1}{c}{Fe5015}&\multicolumn{1}{c}{Mg$_1$} &\multicolumn{1}{c}{Mg$_2$} &\multicolumn{1}{c}{Mg{\it b}} &\multicolumn{1}{c}{Fe5270}&\multicolumn{1}{c}{Fe5335} &\multicolumn{1}{c}{Fe5406} \\

\multicolumn{1}{c}{ }         &\multicolumn{1}{c}{({\AA})}  &\multicolumn{1}{c}{({\AA})} &\multicolumn{1}{c}{({\AA})}  &\multicolumn{1}{c}{({\AA})} &\multicolumn{1}{c}{({\AA})}  &\multicolumn{1}{c}{({\AA})} &\multicolumn{1}{c}{(mag)}&\multicolumn{1}{c}{(mag)}&\multicolumn{1}{c}{({\AA})}&\multicolumn{1}{c}{({\AA})} &\multicolumn{1}{c}{({\AA})}  &\multicolumn{1}{c}{({\AA})}     \\

\hline
GC100	&    1.1303  &    0.6908  &    0.8333  &    0.9859  &    0.4468  &    0.8805  &    0.0100  &    0.0198  &    0.7685  &    0.5803  &    0.5702  &    0.3259  \\ 
GC101	&    1.6954  &    0.7197  &    1.0695  &    1.4733  &    0.6176  &    0.9527  &    0.0125  &    0.0147  &    0.6008  &    0.6563  &    0.6170  &    0.4762  \\ 
GC102	&    1.3257  &    0.6216  &    0.7503  &    0.9542  &    0.5749  &    0.7267  &    0.0110  &    0.0154  &    0.5749  &    0.6034  &    0.6326  &    0.5922  \\ 
GC103	&    1.9956  &    0.9610  &    1.1118  &    1.2570  &    0.5201  &    1.0865  &    0.0114  &    0.0181  &    0.7237  &    0.6221  &    0.5433  &    0.4562  \\ 
GC104	&    1.3665  &    0.5724  &    0.7386  &    0.8269  &    0.5088  &    0.8315  &    0.0091  &    0.0182  &    0.6759  &    0.5374  &    0.5781  &    0.3159  \\ 
GC105	&    0.9879  &    0.5002  &    0.7100  &    0.6995  &    0.6118  &    0.6591  &    0.0080  &    0.0097  &    0.3598  &    0.4706  &    0.4839  &    0.3192  \\ 
GC106	&    0.9924  &    0.4814  &    0.6203  &    0.6667  &    0.4751  &    0.6720  &    0.0076  &    0.0150  &    0.5710  &    0.4319  &    0.4364  &    0.2948  \\ 
GC107	&    0.9105  &    0.4480  &    0.5719  &    0.8968  &    0.4843  &    0.7245  &    0.0077  &    0.0189  &    0.6990  &    0.4699  &    0.4474  &    0.3780  \\ 
GC108	&    1.1974  &    0.7843  &    0.8622  &    0.7959  &    0.4353  &    0.9906  &    0.0106  &    0.0236  &    0.8286  &    0.5325  &    0.5086  &    0.3594  \\ 
GC111	&    0.9826  &    0.4469  &    0.5280  &    0.7158  &    0.4943  &    0.6516  &    0.0065  &    0.0104  &    0.3920  &    0.3709  &    0.4119  &    0.2275  \\ 
\hline
\end{tabular}
\tablecomments{The full table is available in the online version.}
\vspace{0.3cm}
\end{center}
\end{table}

\begin{table}
\begin{center}
\caption{Polynomial coefficients from the orthogonal distance regression fits between MWGCs \citep{kim16} and [Fe/H] by \citet{harris1996}. 
\label{tbl5}}
\begin{tabular}{ccccc}
\hline
\hline
\multicolumn{1}{c}{Index} & 
\multicolumn{1}{c}{a0} & 
\multicolumn{1}{c}{a1} & 
\multicolumn{1}{c}{a2} & 
\multicolumn{1}{c}{Valid Range} \\
\hline
CN$_{1}$ &  $-$0.350 &  10.684 & $-$99.372    & ($-$0.10, 0.12)      \\
CN$_{2}$ &  $-$0.554 &  18.272 & $-$169.425   & ($-$0.08, 0.15)     \\
Ca4227   &  $-$2.646 &  4.463  & $-$1.673     & ($-$0.13, 0.13)     \\
G4300    &  $-$2.645 &  0.580  & $-$0.028     & (0.54, 5.73)     \\
Fe4383   &  $-$2.180 &  0.737  & $-$0.062     & ($-$0.04, 5.14)    \\
Ca4455   &  $-$2.182 &  3.163  & $-$1.129     & ($-$0.42, 1.05)    \\
C$_{2}$4668  &  $-$1.564 &  0.671  & $-$0.078     & ($-$0.44, 4.66)     \\
H$\beta$ &  1.790 &  $-$1.535  &  0.061       & (1.33, 2.80)     \\
Mg$_{1}$ &  $-$1.898 &  35.285 &  $-$166.779  & (0.002, 0.103)    \\
Mg$_{2}$ &  $-$2.526 &  16.363 &  $-$26.171   & (0.03, 0.26)    \\
Mg$b$    &  $-$2.429 &  0.933  &  $-$0.091    & (0.02, 3.84)   \\
Fe5270   &  $-$2.980 &  1.641  &  $-$0.218    & (0.34, 2.85)   \\
Fe5335   &  $-$2.923 &  1.922  &  $-$0.309    & (0.48, 2.51)     \\
Fe5406   &  $-$2.454 &  2.433  &  $-$0.616    & (0.24, 1.74)     \\
H$\delta_{A}$ &  $-$0.612 & $-$0.260  & $-$0.014   & ($-$1.71, 3.89) \\   
H$\delta_{F}$ &  $-$0.175 & $-$0.534  & $-$0.006   & (0.20, 3.01) \\
H$\gamma_{A}$ &  $-$1.444 & $-$0.236  & $-$0.006   & ($-$6.12, 2.52)     \\
H$\gamma_{F}$ &  $-$0.803 & $-$0.400  & $-$0.028   & ($-$1.35, 2.60)     \\
\hline
\end{tabular}
\vspace{0.3cm}
\tablecomments{For a given index $I$, [M/H]$_I$ = a0 + a1$\times$I + a2$\times$I$^2$.}
\end{center}
\end{table}

\clearpage
\begin{table}
\begin{center}
\caption{The [Fe/H] by C98 transformed to our metallicity scale.
\label{tbl6}}
\begin{tabular}{cccccccccc}
\hline
\hline
\colhead{ID} & 
\colhead{[Fe/H]} & 
\colhead{} & 
\colhead{} & 
\colhead{ID} & 
\colhead{[Fe/H]} & 
\colhead{} & 
\colhead{} & 
\colhead{ID} & 
\colhead{[Fe/H]} \\
\hline
5001 & $-$1.046 & & & 579  &  $-$2.088 & & & 1010 & $-$0.213 \\ 
5015 & $-$1.254 & & & 581  &  $-$1.046 & & & 1016 &    0.115 \\ 
5024 & $-$1.551 & & & 588  &  $-$1.211 & & & 1019 & $-$1.299 \\ 
5026 & $-$1.003 & & & 602  &  $-$0.795 & & & 1032 & $-$0.915 \\ 
5028 &    0.115 & & & 611  &  $-$1.748 & & & 1034 & $-$0.915 \\ 
 58  & $-$2.340 & & & 647  &  $-$0.958 & & & 1044 & $-$1.123 \\ 
 66  & $-$1.211 & & & 649  &  $-$1.419 & & & 1093 & $-$0.958 \\ 
107  & $-$0.585 & & & 672  &  $-$1.672 & & & 1155 & $-$0.334 \\ 
141  & $-$0.795 & & & 679  &  $-$0.421 & & & 1157 & $-$1.254 \\ 
176  & $-$0.882 & & & 680  &  $-$1.627 & & & 1158 & $-$1.462 \\ 
177  & $-$0.882 & & & 682  &  $-$1.923 & & & 1180 & $-$1.123 \\ 
186  & $-$1.419 & & & 723  &  $-$2.461 & & & 1240 & $-$0.421 \\ 
248  & $-$0.958 & & & 741  &  $-$0.334 & & & 1247 & $-$0.750 \\ 
279  & $-$0.915 & & & 746  &  $-$0.630 & & & 1293 & $-$0.882 \\ 
280  & $-$0.750 & & & 750  &  $-$1.375 & & & 1309 & $-$0.630 \\ 
290  & $-$0.674 & & & 770  &  $-$0.838 & & & 1336 & $-$1.123 \\ 
307  & $-$1.090 & & & 784  &  $-$0.750 & & & 1344 & $-$0.750 \\ 
314  & $-$0.377 & & & 796  &  $-$1.715 & & & 1351 & $-$0.674 \\ 
321  & $-$1.923 & & & 798  &  $-$1.584 & & & 1370 & $-$0.542 \\ 
323  & $-$2.340 & & & 809  &  $-$1.462 & & & 1382 & $-$2.044 \\ 
324  & $-$0.838 & & & 811  &  $-$2.461 & & & 1449 & $-$1.507 \\ 
330  & $-$0.674 & & & 824  &  $-$0.882 & & & 1463 & $-$1.375 \\ 
348  & $-$1.375 & & & 838  &  $-$1.046 & & & 1479 & $-$1.419 \\ 
350  & $-$1.123 & & & 849  &  $-$0.509 & & & 1481 & $-$1.123 \\ 
376  & $-$1.672 & & & 887  &  $-$1.090 & & & 1504 & $-$2.461 \\ 
417  & $-$0.674 & & & 910  &  $-$2.461 & & & 1514 & $-$0.301 \\ 
418  & $-$1.551 & & & 928  &  $-$1.507 & & & 1538 & $-$0.674 \\ 
421  & $-$0.465 & & & 941  &  $-$1.748 & & & 1548 & $-$0.915 \\ 
423  & $-$0.377 & & & 952  &  $-$1.419 & & & 1565 & $-$1.584 \\ 
442  & $-$1.299 & & & 965  &  $-$0.717 & & & 1594 & $-$1.211 \\ 
453  & $-$1.090 & & & 968  &  $-$1.254 & & & 1615 & $-$1.046 \\ 
490  &    0.115 & & & 970  &  $-$0.795 & & & 1617 & $-$1.046 \\ 
491  & $-$1.715 & & & 991  &  $-$1.419 & & & 1631 & $-$1.342 \\ 
526  & $-$0.257 & & & 007  &  $-$0.630 & & & 1664 & $-$0.838 \\ 

\hline
\end{tabular}
\vspace{0.3cm}
\end{center}
\end{table}

\clearpage
\begin{table}
\begin{center}
\caption{The result of the GMM analysis for the metallicity distributions shown in the right panel of Figure \ref{M87_14}. \label{tbl7}}
\begin{tabular}{cc}
\hline
\hline
[Fe/H]$_{\rm This}$\,$_{\rm study}$ & [Fe/H]$_{\rm Combined}$ \\
\hline
\multicolumn{2}{c}{{\it P}-value$^{a}$} \\
$P(\chi^2)$, $P(DD)$, $P(kurt)$ & $P(\chi^2)$, $P(DD)$, $P(kurt)$ \\
\hline
0.120,\,\,\,0.406,\,\,\,0.192 & 0.246,\,\,\,0.652,\,\,\,0.358  \\
\hline		    
\multicolumn{2}{c}{Number Ratio$^{b}$} \\	  
\hline
0.371\,:\,0.629 & 0.371\,:\,0.629 \\
\hline
\multicolumn{2}{c}{$\mu^{c}_{b}$, $\mu^{c}_{r}$} \\	  
\multicolumn{2}{c}{($\sigma^{c}_{b}$, $\sigma^{c}_{r}$)} \\	  
\hline
$-$1.607, $-$0.666 & $-$1.542, $-$0.805 \\
(0.395, 0.346) & (0.506, 0.421) \\
\hline
\end{tabular}
\vspace{0.3cm}
\tablecomments{
$^{a}$ The resulting significance of the GMM test expressed as a {\it P}-value. \\
$^{b}$ The number ratios between the blue and red GC groups.\\ 
$^{c}$ The mean and the standard deviations of the blue and red GC groups.\\
} 
\end{center}
\end{table}
\begin{table}
\begin{center}
\caption{The result of the GMM analysis for the color distributions shown in the fourth row of Figure \ref{M87_15}. \label{tbl8}}
\begin{tabular}{cc}
\hline
\hline
$V-I$ & $M-T2$ \\
\hline
\multicolumn{2}{c}{{\it P}-value$^{a}$} \\
$P(\chi^2)$, $P(DD)$, $P(kurt)$ & $P(\chi^2)$, $P(DD)$, $P(kurt)$ \\
\hline
0.001,\,\,\,0.167,\,\,\,0.003 & 0.001,\,\,\,0.094,\,\,\,0.001  \\
\hline		    
\multicolumn{2}{c}{Number Ratio$^{b}$} \\	  
\hline
0.602\,:\,0.398 & 0.554\,:\,0.446 \\
\hline
\multicolumn{2}{c}{$\mu^{c}_{b}$, $\mu^{c}_{r}$} \\	  
\multicolumn{2}{c}{($\sigma^{c}_{b}$, $\sigma^{c}_{r}$)} \\	  
\hline
1.021, 1.219 & 1.224, 1.456 \\
(0.058, 0.062) & (0.062, 0.063) \\
\hline
\end{tabular}
\vspace{0.3cm}
\tablecomments{
$^{a}$ The resulting significance of the GMM test expressed as a {\it P}-value. \\
$^{b}$ The number ratios between the blue and red GC groups.\\ 
$^{c}$ The mean and the standard deviations of the blue and red GC groups.\\
}
\end{center}
\end{table}

\end{document}